\documentclass[lettersize,journal]{IEEEtran}

\usepackage{amsmath,amsfonts}
\usepackage{algorithmic}
\usepackage{booktabs}
\usepackage{algorithm}
\usepackage{array}
\usepackage{siunitx}
\usepackage[caption=false,font=normalsize,labelfont=sf,textfont=sf]{subfig}
\usepackage{textcomp}
\usepackage{url}
\usepackage{verbatim}
\usepackage{amssymb}
\usepackage{mathtools}
\usepackage{graphicx}
\usepackage{epstopdf}
\usepackage{multirow}
\usepackage{caption}        
\usepackage{subcaption}
\usepackage{epstopdf}
\usepackage[utf8]{inputenc}
\usepackage{graphicx}
\usepackage{subcaption}
\usepackage{float}
\usepackage{caption} 
\usepackage{authblk}
\usepackage{bm}
\usepackage[font=small]{caption}
\usepackage{xcolor}
\usepackage{hyperref}
\usepackage{amsmath}
\usepackage{makecell}
\usepackage{cite}

\abovedisplayshortskip
\belowdisplayshortskip

\DeclareGraphicsExtensions{.jpg,.png,.pdf}
\begin{document}

\title{Enhancing Energy and Spectral Efficiency in IoT-Cellular Networks via Active SIM-Equipped LEO Satellites}

\author{Rahman Saadat Yeganeh\,$^1$, Hamid Behroozi\,$^{1}$, Mohammad Javad Omidi\,$^2$, Mohammad Robat Mili\,$^3$, 

Eduard A. Jorswieck\,$^4$,~\IEEEmembership{Fellow, IEEE},  Symeon Chatzinotas\,$^{5},$~\IEEEmembership{Fellow, IEEE},

\thanks{$^1$Department of Electrical Engineering Sharif University of Technology Tehran, Iran (Emails: rahman.saadat@sharif.edu, behroozi@sharif.edu)}
\thanks{$^2$Department of Electronics and Communication Engineering
Kuwait College of Science and Technology
Doha, Kuwait (Email: omidi@iut.ac.ir).}
\thanks{$^3$Department of Computer and Engineering, PIAIS, IRAN (Email: mohammad.robatmili@gmail.com).}
\thanks{$^4$Institute for Communications Technology
Technische Universität Braunschweig
Braunschweig, Germany (Email: e.jorswieck@tu-braunschweig.de)}
\thanks{$^5$Interdisciplinary Centre for Security, Reliability and Trust, University of Luxembourg
(Email: Symeon.Chatzinotas@uni.lu).}}

\maketitle
\begin{abstract}

This paper investigates a low Earth orbit (LEO) satellite communication system enhanced by an active stacked intelligent metasurface (ASIM), mounted on the backplate of the satellite's solar panels to efficiently utilize limited onboard space and reduce the main satellite power amplifier requirements. The system serves multiple ground users via rate-splitting multiple access (RSMA) and IoT devices through a symbiotic radio network. Multi-layer sequential processing in the ASIM improves effective channel gains and suppresses inter-user interference, outperforming active RIS and beyond-diagonal RIS designs.
Three optimization approaches are evaluated: block coordinate descent with successive convex approximation (BCD-SCA), model-assisted multi-agent constraint soft actor-critic (MA-CSAC), and multi-constraint proximal policy optimization (MCPPO). Simulation results show that BCD-SCA converges fast and stably in convex scenarios without learning, MCPPO achieves rapid initial convergence with moderate stability, and MA-CSAC attains the highest long-term spectral and energy efficiency in large-scale networks. Energy–spectral efficiency trade-offs are analyzed for different ASIM elements, satellite antennas, and transmit power. Overall, the study demonstrates that integrating multi-layer ASIM with suitable optimization algorithms offers a scalable, energy-efficient, and high-performance solution for next-generation LEO satellite communications.
\end{abstract}


\section{Background}
\label{sec:background}

6G wireless networks require efficient solutions to connect billions of IoT devices, cellular users, and other wireless technologies. IoT is crucial for applications such as smart transportation, homes, grids, and agriculture \cite{you2021towards}. However, IoT deployment faces challenges in energy efficiency and spectrum scarcity \cite{r1,r3,r28}, as frequency resources are limited and device batteries incur high maintenance costs \cite{zhang20196g}.

Symbiotic Radio (SR) has emerged as a promising approach, building on ambient backscatter communication \cite{r19,r26}. SR networks are categorized into parasitic SR (PSR) and commensal SR (CSR) \cite{8907447,r26}. PSR supports high data rates but suffers from interference, requiring complex cancellation. CSR is better suited for low-data-rate IoT networks, reducing interference through joint decoding and transmit collaboration \cite{r22,r25}.

Increasing network frequencies and growing IoT populations can create coverage blind spots. Satellite systems, particularly LEO satellites, can bridge these gaps, providing low-latency, real-time data collection, and enhanced IoT integration for smart cities \cite{you2024ubiquitous,gongora2022link}. However, satellites face weak link budgets and signal attenuation, especially at higher frequencies.

Reconfigurable Intelligent Surfaces (RIS) offer a cost-effective solution to mitigate signal attenuation, enhancing spectrum and energy efficiency \cite{wu2021intelligent,an2021low}. Traditional single-layer RIS designs are limited in beamforming flexibility. The SIM improves upon this by integrating multiple metasurfaces with antenna arrays, enabling better transmit precoding and receiver combining for higher capacity and energy efficiency \cite{an2023stacked}.
\textcolor{black}{
ASIM is particularly attractive for LEO systems because it combines active multi-layer control, limited-power amplification, and on-board energy harvesting to cope with LEO-specific impairments (large path loss, rapid Doppler, intermittent visibility, and tight link budgets). Recent layered-metasurface studies illustrate complementary design principles. Dual-polarized stacked metasurface transceivers with rate-splitting demonstrate how layered metasurfaces can assist RSMA by mapping common and private messages onto independent polarizations, thereby introducing additional DoFs in power, spatial, and polarization domains \cite{sun2025dual}; active–passive cascaded RIS architectures enable hybrid-layer designs that realize jamming nulling and signal enhancement with tractable semi-closed-form optimization for the active coefficients \cite{sun2023active}; and multi-layer refracting RIS receivers support simultaneous information-and-energy transfer in long-range non-terrestrial links while mitigating severe large-scale fading \cite{an2024exploiting}. Building on these principles, our work integrates a multi-layer active SIM on the satellite backplate and uniquely combines RSMA for terrestrial-user multiplexing with commensal backscatter for IoT, governed by a unified optimization and DRL control framework that explicitly accounts for on-board energy harvesting and LEO dynamics.
}

\subsection{Related Works}
Several studies have explored the integration of RIS with satellite communication systems, particularly in LEO networks, to address challenges like path loss, energy efficiency, and interference. In \cite{tekbiyik2021energy}, RIS-assisted satellite networks were used to mitigate path loss caused by long transmission distances. Research in \cite{khan2023ris} optimized energy efficiency by jointly optimizing transmit power and passive beamforming. Another study \cite{lv2024energy} focused on maximizing energy efficiency in RIS-enhanced NOMA satellite networks by optimizing transmit power and phase shifts.

In \cite{10879566}, active simultaneously transmitting and reflecting RIS were employed to enhance 6G cellular networks. A CSR network enables communication between passive IoT and active users, with massive MIMO antennas relaying signals via NOMA and SIC. The optimization problem was addressed using deep reinforcement learning.

In \cite{asif2024transmissive}, transmissive RIS was used to maximize the sum rate of LEO satellite networks. Khan et al. \cite{khan2024cr} proposed a cognitive radio approach for maximizing the sum rate in two-tier satellite networks using RIS. 
Yeganeh et al. \cite{yeganeh2025energy} proposed a non-terrestrial communication system integrating RSMA with BD-ARIS on a UAV under LEO satellite coverage, optimizing beamforming, UAV positioning, and power allocation using DRL algorithms to maximize energy efficiency. TRPO outperformed other algorithms in energy efficiency and sum rate, while A3C showed instability.

Recent developments in SIM have shown promise. In \cite{lin2024stacked}, SIM reduced computational load and processing delay in LEO satellite systems by enabling multiuser beamforming directly in the electromagnetic domain. Further studies, such as \cite{11003236} and \cite{10643881}, highlight SIM's ability to enhance satellite-to-ground communication in extreme environments, improving signal transmission and maximizing SNR.

\subsection{Contribution}
\label{sec:contribution}

This paper presents a novel satellite communication system that employs ASIM on the satellite and leverages SR to enhance connectivity for IoT devices. While ASIM overcomes the limitations of traditional RIS by using stacked metasurfaces to enhance beamforming and signal shaping, SR offers a unique solution for IoT networks by allowing direct communication between IoT devices without the need for frequency allocation or dedicated infrastructure. By leveraging SR, IoT devices can communicate efficiently with minimal energy consumption, addressing spectrum scarcity and energy efficiency challenges. The combination of ASIM and SR in our system enhances both the flexibility of satellite communications and the scalability of IoT networks, paving the way for a more connected and energy-efficient future in 6G systems. \textcolor{black}{
In addition, the adoption of ASIM in our LEO system is motivated by its ability to overcome the severe path-loss, Doppler dynamics, and link-budget constraints of LEO communications, offering higher beamforming flexibility and improved reliability compared with conventional passive RIS or relay solutions.}
The main contributions of this work are summarized as follows:
\begin{itemize}
    \item \textbf{System Design:} We introduce an innovative satellite communication system utilizing active SIMs integrated on the backplate of the satellite's solar panels. This design optimizes space utilization, allowing for compact, efficient, and adaptive beamforming to enhance coverage and signal quality. The use of active SIMs enables dynamic beamforming and precise signal control on the Earth's surface, improving communication performance across varying user locations.

    \item \textbf{Power Amplifier Optimization:} We introduce a novel power amplifier architecture that splits the power amplification into two parts: one for the main satellite system and one for the active ASIM antenna. This approach reduces the overall power consumption and system costs while maintaining high efficiency, overcoming the need for large and energy-intensive power amplifiers typically used in satellite communication systems.

    \item \textbf{Non-Convex Optimization Problem and Solution Framework:} We formulate a non-convex optimization problem to jointly optimize power allocation, satellite beamforming vectors, and ASIM reflection coefficients. The goal is to maximize both energy and spectral efficiency while meeting Quality of Service (QoS) constraints for users. To solve this problem, we employ the BCD-SCA technique and leverage DRL algorithms MA-CSAC and MCPPO to address the complex system constraints and improve performance in real-time.

    \item \textbf{Simulation Results and Performance Analysis:} Extensive simulations demonstrate that the MA-CSAC algorithm achieves superior long-term performance in terms of spectral and energy efficiency, particularly in large-scale networks. We also show that BCD-SCA provides rapid convergence and stability in convex problems, while MCPPO offers faster initial convergence with some trade-offs in stability. These results highlight the effectiveness of the proposed system in optimizing both energy and spectral efficiency.

    \item \textbf{Scalable and Efficient System Design:} We analyze the scalability of the proposed system by examining the trade-offs between SE and EE as the network size grows. Our results show that the combination of active SIMs, power amplifier optimization, and advanced optimization algorithms effectively supports large-scale satellite communication networks. MA-CSAC excels in scalability, maintaining high EE, while BCD-SCA is more suited for energy-constrained environments, offering a balanced solution for next-generation systems.

\end{itemize}

\textcolor{black}{The remainder of this paper is organized as follows. Section II describes the system model. Section III formulates the optimization problem. Section IV proposes a mathematical optimization framework to solve the problem. Section V develops a DRL-based solution. Section VI presents simulation results, and Section VII concludes the paper.}

\textcolor{black}{
\textbf{\emph{Notations}}: 
$\mathbf{A}^T$, $\mathbf{A}^H$, and $\mathrm{Tr}(\mathbf{A})$ are the transpose, conjugate transpose, and trace of $\mathbf{A}$. 
$\|\mathbf{A}\|_F$ and $\|\mathbf{a}\|$ represent the Frobenius norm and Euclidean norm, respectively. 
$\operatorname{diag}(\cdot)$ creates a diagonal matrix. 
}

\begin{table}
	\centering
	    \caption{List of abbreviations.}
	\begin{tabular}{|c|c|}

		\hline  BCD & Block Coordinate Descent   \\
				\hline  CSR & Commensal Symbiotic Radio\\ 
				\hline  DRL & Deep Reinforcement Learning\\ 
				\hline  IoT & Internet of Things\\
		\hline  MA-CSAC & Multi Agent Constraint Soft Actor Critic\\ 
		\hline MCPPO & Multi Constraint Proximal Policy Optimization\\ 
		\hline  NOMA & Non-Orthogonal Multiple Access   \\
\hline  LoS &  Line of Sight  \\
		\hline  LEO & Low Earth orbit\\
		\hline CSI & Channel State Information\\
		\hline P.A. & Power Amplifier\\
		\hline  QoS & Quality of Service\\ 
		\hline  RIS & Reconfigurable Intelligent Surfaces\\
		\hline RSMA & Rate Splitting Multiple Access\\
		\hline ASIM & Active Stacked Intelligent Metasurface \\
		\hline SINR & Signal-to-Interference-plus-Noise Ratio\\
		\hline  SR & Symbiotic Radio\\  
		\hline  SBD & Symbiotic Backscatter Devices\\ 
		\hline  SCA & Successive Convex Approximation\\ 
		\hline SIC & Successive Interference Cancellation\\

		\hline 

	\end{tabular} 
\end{table}

\section{System Model}
\label{sec:system_model}

\begin{figure}
\centering
\includegraphics[width=8cm]{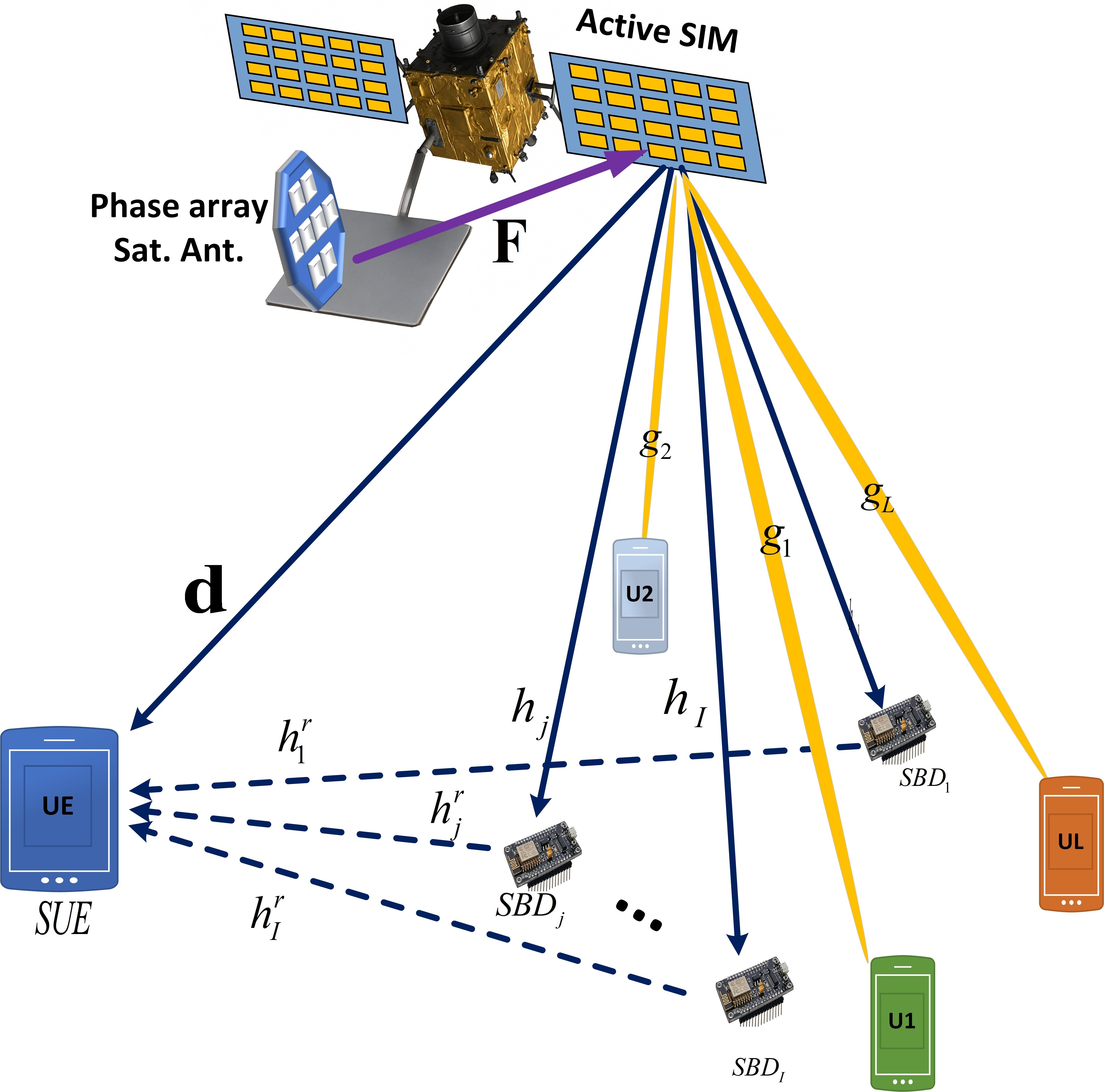}
\caption{System model of an ASIM-assisted LEO satellite for enhancing non-terrestrial communication.}
\label{fig:sat_ris}
\end{figure}

As illustrated in Fig.~\ref{fig:sat_ris}, the considered system model involves a satellite operating in LEO, designed to deliver communication services to multiple terrestrial users. The satellite is equipped with a transmitter of power \( P^{\text{Sat}} \) and a phased-array antenna consisting of \( N \) active elements. These antenna elements steer the transmitted signals toward a SIM, which is integrated on the backplate behind the satellite’s solar panels to optimize space utilization. The SIM comprises \( Q \) stacked reconfigurable layers, each containing \( M \) active elements, capable of performing reflection, amplification, and programmable inter-layer signal propagation. Such a configuration allows the ASIM to dynamically reshape beams and enhance coverage, improving signal quality across diverse ground locations.

The system serves \( L \) legitimate terrestrial users, and their access to radio resources is managed using RSMA, which efficiently mitigates multi-user interference and enhances spectral efficiency. In addition, \( I \) passive IoT devices operate in the same frequency band by modulating their information onto the incident satellite carrier signals, forming a symbiotic radio network. In this paradigm, active users and passive IoT devices coexist and share the spectrum cooperatively, enabling mutual data exchange and improved resource utilization.

The channel model includes the satellite-to-ASIM link and the SIM-to-ground-user links, taking into account path loss, fading, and thermal noise. The goal is to jointly optimize power allocation, satellite beamforming vectors, and ASIM reflection coefficients to maximize energy and spectral efficiency while satisfying the users' QoS constraints.

\subsection{RSMA System and Active SIM-Enhanced Satellite Network Model}

The considered system features a LEO satellite employing RSMA to simultaneously serve \(L\) legitimate terrestrial users. The satellite is equipped with an \(N\)-element phased-array antenna and integrates an ASIM, composed of \(M\) tunable meta-atoms arranged across \(Q\) programmable metasurfaces. Each meta-atom, a fundamental building block of the ASIM, is constructed using advanced reconfigurable metamaterials, enabling real-time control over the amplitude, phase, and polarization of incident electromagnetic signals. To maximize structural efficiency, the ASIM is mounted on the Earth-facing backplate behind the satellite's solar panels, thereby enhancing downlink communication performance without occupying additional space.

The ASIM's metasurface architecture allows for flexible three-dimensional wave manipulation, including dynamic beamforming, spatial filtering, directional gain control, and interference suppression. These reconfigurable capabilities adapt to varying network conditions and user locations, thereby improving coverage, signal quality, and overall SE.

The satellite transmits a baseband signal vector \( \mathbf{x}(t) \in \mathbb{C}^{N \times 1} \), consisting of a common stream \( \mathbf{x}_c(t) \) and \(L\) private streams \( \mathbf{x}_l(t),\forall l \in \psi, \psi = \{1, 2, \dots, L\}\). The common stream is intended for all users, while each private stream is intended for a specific user. All data symbols are assumed to be mutually independent and normalized to unit power, facilitating efficient power allocation across streams.

The satellite's precoding matrix is denoted as
\(
\mathbf{W} = [\mathbf{w}_c, \mathbf{w}_1, \ldots, \mathbf{w}_L] \in \mathbb{C}^{N \times (L+1)},
\)
\textcolor{black}{where \(\mathbf{w}_c,\mathbf{w}_l \in \mathbb{C}^{N \times 1}\) for \(l=1,\ldots,L\)}. The transmitted signal is expressed as:
\begin{equation}
\mathbf{x}(t) = \sqrt{\sigma_c P^{\text{sat}}} \mathbf{w}_c x_c(t) + \sum_{l=1}^L \sqrt{\sigma_l P^{\text{sat}}} \mathbf{w}_l x_l(t),
\label{eq:transmitted_signal}
\end{equation}where \(\sigma_c, \sigma_l \geq 0\) are the power allocation coefficients satisfying \(\sigma_c + \sum_{l=1}^L \sigma_l = 1\). This allocation ensures a balance between maximizing spectral efficiency and minimizing energy consumption over the satellite-user links.

The transmitted signal first propagates through the satellite-to-ASIM channel \( \mathbf{F} \in \mathbb{C}^{M \times N} \), representing the wireless link between the phased-array antenna and the ASIM. 
Each metasurface layer \( q \in \{1, \ldots, Q\} \) within the ASIM includes a diagonal active tuning matrix:
\begin{equation}
\mathbf{\Phi}^{(q)} = \operatorname{diag}\bigl(\Phi_1^{(q)}, \ldots, \Phi_M^{(q)}\bigr) \in \mathbb{C}^{M \times M},
\label{eq:phi_k}
\end{equation}
\textcolor{black}{where the complex gain of the \(m\)-th meta-atom is expressed as \(\Phi_m^{(q)} = |\Phi_m^{(q)}| e^{j\angle\Phi_m^{(q)}}\). Unlike passive metasurfaces, the active elements allow amplitude values \(|\Phi_m^{(q)}|\) greater than unity, thereby providing signal amplification.} Through the joint tuning of \(|\Phi_m^{(q)}|\) and \(\angle\Phi_m^{(q)}\), each metasurface layer performs coordinated amplitude and phase adjustment, enabling three-dimensional beam shaping, gain control, and interference suppression across the stacked ASIM structure.

The inter-layer coupling and propagation effects between metasurfaces \(q-1\) and \(q\) are modeled by full-rank channel matrices \(\mathbf{H}^{(q)} \in \mathbb{C}^{M \times M}\), which capture near-field interactions. The overall transfer matrix \textcolor{black}{ \(\mathbf{T} \in \mathbb{C}^{M \times M}\)} of the ASIM is:
\begin{equation}
\mathbf{T} = \mathbf{\Phi}^{(Q)} \mathbf{H}^{(Q)} \cdots \mathbf{\Phi}^{(1)} \mathbf{H}^{(1)},
\label{eq:sim_transfer_prod}
\end{equation}

\textcolor{black}{where \(\mathbf{H}^{(q)}\) models the propagation channel from the output of layer \(q-1\) to the input of layer \(q\). This product ordering reflects the forward propagation of the signal as it traverses the ASIM from the first metasurface layer to the last. Since the geometry and spacing of the metasurface layers are fixed by design, the inter-layer coupling matrices \(\mathbf{H}^{(q)}\) are deterministic and assumed to be perfectly known at the ASIM controller.} The inter-layer channel matrices \(\mathbf{H}^{(q)}\) capture the near-field electromagnetic coupling and wave propagation between adjacent metasurface layers, including mutual coupling effects, spatial dispersion, and coordinated amplitude–phase interactions enabled by the stacked ASIM architecture, and can be effectively characterized using compact matrix representations without resorting to full-wave electromagnetic simulations.

The output signal from the ASIM is modeled as:
\begin{equation}
\mathbf{R}^{\text{Out}}(t) = \mathbf{T} \mathbf{F} \mathbf{x}(t) + \mathbf{n}_{\text{SIM}}(t),
\label{eq:sim_output_signal}
\end{equation}
where \(\mathbf{n}_{\text{SIM}}(t) \sim \mathcal{CN}(\mathbf{0}_M, \sigma_{\text{SIM}}^2 \mathbf{I}_M)\) accounts for thermal and circuit noise induced by the active tuning elements within the ASIM. \textcolor{black}{This term represents the effective aggregate noise at the ASIM output, capturing the cumulative contribution of per-layer noise sources propagated through the subsequent metasurface layers.}

The ASIM-to-user-\(l\) downlink channel is denoted by \(\mathbf{g}_l \in \mathbb{C}^{M \times 1}\), which is imperfectly estimated as \(\hat{\mathbf{g}}_l = \mathbf{g}_l + \mathbf{e}_l\), with the estimation error \(\mathbf{e}_l \sim \mathcal{CN}(\mathbf{0}_M, \sigma_{e_l}^2 \mathbf{I}_M)\) accounting for Doppler shifts, feedback delay, and measurement imperfections.
\textcolor{black}{
The ASIM-to-user channels incorporate large-scale path loss determined by the satellite altitude and elevation angle, small-scale fading and multipath effects due to ground scattering, as well as Doppler-induced impairments arising from the high orbital velocity of the LEO satellite, which are implicitly reflected in the adopted channel estimation error model.
}

The received signal at user \(l\) is then:
\begin{equation}
y_{U_l}(t) = \mathbf{g}_l^H \mathbf{R}^{\text{Out}}(t) + n_{U_l}(t),
\label{eq:user_received_signal}
\end{equation}
where \(n_{U_l}(t) \sim \mathcal{CN}(0, \sigma_{U_l}^2)\) represents additive receiver noise.

Additionally, the system supports \(I\) passive IoT devices, which utilize symbiotic backscatter communication. These devices modulate their information onto the reflection coefficient of the incident signal, enabling spectrum sharing with the RSMA users without orthogonal resource allocation. This design achieves ultra-low-power transmission, ideal for energy-constrained IoT scenarios.

\textcolor{black}{
We assume perfect CSI for the satellite-to-ASIM link \( \mathbf{F} \), owing to its deterministic geometry and static configuration. In contrast, the ASIM-to-user channels are subject to estimation errors (imperfect CSI), and thus the proposed algorithms perform robust joint optimization of the satellite precoding vectors \( \mathbf{w}_c, \mathbf{w}_l \) and the SIM tuning matrices \( \{\mathbf{\Phi}^{(q)}\}_{q=1}^Q \) to ensure user-specific QoS under channel uncertainty. This design explicitly accounts for CSI imperfections in the optimization process, demonstrating that the algorithms remain suitable and effective even when the channel estimates are not exact.
}

\subsection{Symbiotic Radio Network Model}

As discussed, enabling full communication among passive IoT users is a crucial requirement in next-generation communication networks. In the system model presented in this paper, the ground IoT network structure is modeled as a CSR to simulate a realistic operational environment. In SR systems, it is assumed that each backscatter symbol period spans $K$ satellite symbol periods, meaning the symbol duration for backscatter transmissions is $K$ times that of a satellite symbol. For $K \gg 1$, the satellite transmits $k = \{1,2,\dots,K\}$ symbols per backscatter symbol period, denoted $\mathbf{x}_k(t)$, while each SBD modulates a single information symbol $s(t)$ onto the received signal and forwards it to the target SUE. Synchronization between $\mathbf{x}_k(t)$ and $s(t)$ is necessary to prevent spectral expansion when $K = 1$; however, as $K$ increases, the impact of asynchronism becomes negligible \cite{guo2019cognitive}. \textcolor{black}{We adopt a discrete time-slot model wherein each slot $t$ corresponds to the duration of one backscatter symbol period, i.e., $K$ satellite symbol intervals, and all channel states, precoding vectors, and SIM configurations remain constant within the slot.} Accordingly, in Eq. \ref{eq:sim_output_signal}, ${\mathbf{x}_c}(t)$ and ${\mathbf{x}_l}(t)$ are updated to $\mathbf{x}_c^k(t)$ and $\mathbf{x}_l^k(t)$ to reflect individual satellite symbol transmissions.

\begin{figure}
\includegraphics[width=8.7cm]{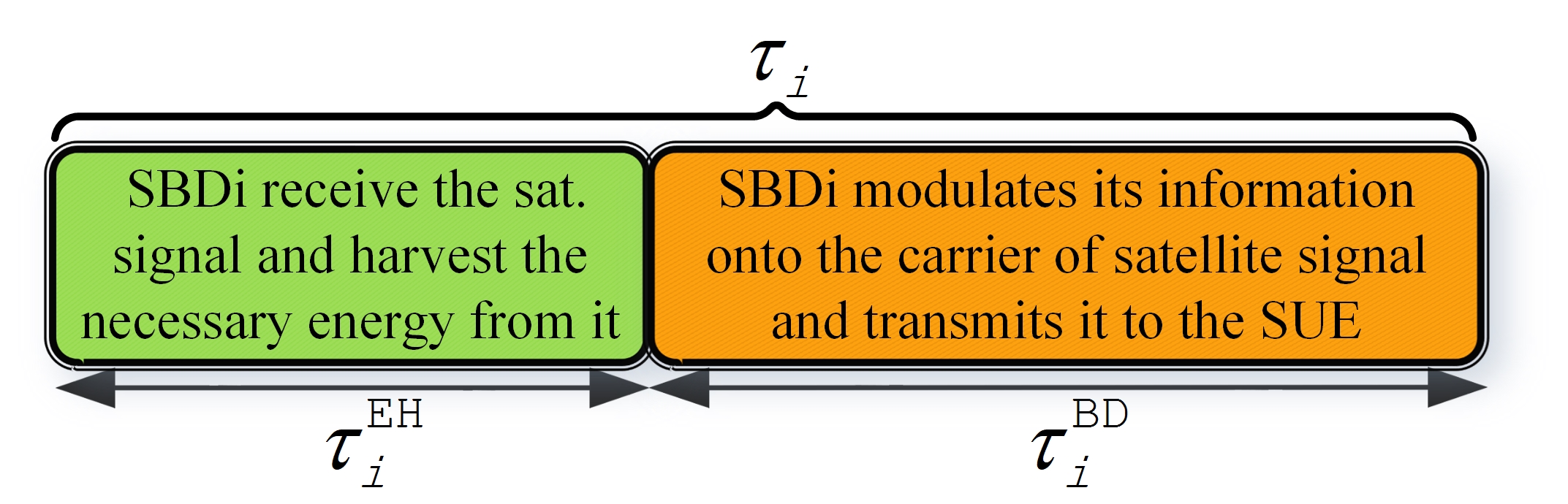}
\caption{The time division duplexing in the CSR network.} 
\label{fig:sat_tdd}
\end{figure}
According to Fig. \ref{fig:sat_tdd}, each SBD equipped with a single antenna harvests energy during the period \( \tau_i^{\mathrm{EH}} \) and transmits its information during \( \tau_i^{\mathrm{BD}} \), with \( \tau_i^{\mathrm{EH}} + \tau_i^{\mathrm{BD}} = 1 \). These processes are defined over a normalized transmission frame of unit duration, so that \( \tau_i^{\mathrm{EH}} \) and \( \tau_i^{\mathrm{BD}} \) represent the respective time fractions. Under this normalization, the harvested energy is obtained directly from the received signal power integrated over \( \tau_i^{\mathrm{EH}} \), ensuring dimensional consistency. \textcolor{black}{Hence, the term \( \tau_i^{\mathrm{EH}}|\mathbf{h}_i^H \mathbf{R}^{\mathrm{Out}}(t)|^2 \) yields the harvested energy up to the efficiency factor \( \Gamma \).}

Each SBD harvests energy up to the maximum level required by its power source, based on the amount of information it needs to transmit. Following information transmission and energy depletion, it returns to the energy harvesting phase. This cycle continues periodically throughout the network \cite{yeganeh2023multi}.

With the set \( \upsilon = \left\{ 1, 2, \dots, I \right\} \) defined as all passive IoT users in the CSR network (SBDs), the received signal by \( \text{SBD}_i \) is given by 
\(
y_{\rm{SBD_i}}(t) = {\mathbf{h}}_i^H {\mathbf{R}}^{\rm{Out}}(t) + n_{\rm{SBD_i}}(t), \forall i \in \upsilon.
\) 
This device first harvests energy from the received signal as follows:
\begin{equation}
\label{eq:EH}
{\varepsilon _{{\rm{SB}}{{\rm{D}}_i}}} \le {\Gamma}\tau _i^{{\rm{EH}}}|{\bf{h}}_i^H{\bf{R}}^{Out}\left( t \right){|^2},{\kern 1pt} {\kern 1pt} {\kern 1pt} {\kern 1pt} {\kern 1pt} {\kern 1pt} {\kern 1pt} \forall i \in \upsilon
 \end{equation}
In Eq. \ref{eq:EH}, \( 0 \leq {\Gamma} \leq 1 \) represents the energy conversion efficiency from the received waves. It should be noted that since SBDs do not have any power-consuming active components, \( n_{\rm{SBD_i}}(t)\), which is the complex Gaussian noise at the SBD's antenna, is very small and can be disregarded \cite{r41}. The \( \text{SBD}_i \) modulates its information signal \( s_i(t) \) onto the received satellite carrier signal using the harvested energy and reflects it towards the destination (SUE) with a reflection coefficient \(\eta _i\). It is worth noting that the SUE is itself an active receiver within the intended legal network in this structure. The received signal at this user, after passing through the channel \(h_i^r\), is given by:
\begin{equation}
\label{eq:y_SUE}
\begin{gathered}
  y_{\rm{SUE}}^k\left( t \right) = \sum\limits_{i = 1}^I {\sqrt {{\eta _i}} h_i^r{{\bf{h}}_i^H}{\mathbf{R}}^{Out}\left( t \right){s_i}\left( t \right)}  +  {\mathbf{d}^{H}\mathbf{R}}^{\rm{Out}}\left( t \right)\hfill \\ 
  \,\,\,\,\,\,\,\,\,\,\,\,\,\,\,\,\,\,\,\,\,\,\,\,+ n_{\rm{SUE}}^k\left( t \right), \,\,\,\,\,\forall i \in \upsilon , k = \left\{ {1,2,...,K} \right\}  \hfill \\ 
\end{gathered} 
 \end{equation}
Here, \( n_{\rm{SUE}}^k(t) \sim \mathcal{CN}(0, \sigma_{\rm{SUE}}^{2,k}) \) is associated with the AWGN at the SUE.

\subsection{Data Rate Model}

\subsubsection{RSMA Network}

In the RSMA framework, each user first decodes the common stream \( x_c \) during time slot \( t \). Upon successfully decoding and subtracting this common component from the received signal, each user proceeds to decode its own private stream \( x_l \). The SINR for decoding the common stream at user \( l \), assuming no denoising is performed at the ASIM, is given by:
\textcolor{black}{
\begin{equation}
\label{eq:SINR_common_final}
\gamma_{c,l}(t) =
\frac{
    \sigma_c P^{\mathrm{sat}}
    \left|  \mathbf{U}_c \right|^2}
    {
    \sum_{l=1}^{L}
    \sigma_l P^{\mathrm{sat}}
    \left|{\mathbf{U}_l}\right |^2
    + 
    {\left\| \mathbf{g}_l^H\mathbf{T} \right\|^2}\sigma_{\mathrm{SIM}}^2 
    + N_0
},
\end{equation}
}

where \( \mathbf{U}_c \triangleq \mathbf{g}_l^H \mathbf{TF} \mathbf{w}_c \) and \( \mathbf{U}_l \triangleq \mathbf{g}_l^H \mathbf{TF} \mathbf{w}_l \) denote the effective combined channels for the common and private signals, respectively,
and \( N_0 \) is the receiver noise power spectral density (W/Hz).

According to Shannon’s capacity formula, the achievable \textcolor{black}{spectral efficiency} for the common stream at user \( l \) (in bps/Hz) is expressed as: 
\begin{equation}
\label{eq:R_c_psd}
R_{c,l}(t) =  \log_2 \!\left( 1 + \gamma_{c,l}(t) \right), \quad \forall l \in \psi.
\end{equation}

To ensure reliable decoding of the common stream by all users, the allocated common rate \( R_c(t) \) must not exceed the minimum achievable rate among all users, i.e.,
\[
R_c(t) \leq \min \{ R_{c,1}(t), R_{c,2}(t), \dots, R_{c,L}(t) \}.
\]
Moreover, since the common stream is intended for all users, it is partitioned into user-specific portions such that:
\[
\sum_{l=1}^{L} C_l(t) = R_c(t),
\]
where \( C_l(t) \) denotes the rate share of the common stream intended for user \( U_l \) \cite{xu2021rate}. \textcolor{black}{Hence, the common-rate allocation rule under RSMA }can be summarized as:
\begin{equation}
\label{eq:C_l}
R_c(t) = \sum_{l=1}^{L} C_l(t) \leq \min_{l} \{ R_{c,l}(t) \}, \quad \forall l \in \psi.
\end{equation}

After removing the common stream via SIC, each user proceeds to decode its private stream. The SINR for decoding the private stream of user \( l \) is given by:
\textcolor{black}{\begin{equation}
\label{eq:SINR_p_psd}
\gamma_{p,l}(t) =
\frac{\sigma_l P^{\mathrm{sat}} \left| \mathbf{U}_l \right|^2}
{\displaystyle \sum_{\ell \in \psi, \ell \neq l}     \sigma_\ell P^{\mathrm{sat}}
    \left|{\mathbf{U}_\ell}\right |^2
    + 
    {\left\| \mathbf{g}_l^H \mathbf{T} \right\|^2}\sigma_{\mathrm{SIM}}^2 
    + N_0},
\end{equation}
}where \( \mathbf{U}_\ell \triangleq \mathbf{g}_l^H \mathbf{TF} \mathbf{w}_\ell \), and the denominator includes the interference from other users’ private streams as well as the ASIM-induced distortion noise.  
\textcolor{black}{
The corresponding achievable \textcolor{black}{spectral efficiency} for decoding the private stream \( x_l \) is then:
\begin{equation}
\label{eq:R_p_psd}
R_{p,l}(t) =  \log_2 \!\left( 1 + \gamma_{p,l}(t) \right), \quad \forall l \in \psi.
\end{equation}}
Finally, the total achievable sum-rate for the RSMA-based satellite communication system at time slot \( t \) is given by:
\begin{equation}
\label{eq:R_sum}
R_{\text{sum}}(t) = \sum_{l = 1}^L \left( C_l(t) + R_{p,l}(t) \right), [\text{bps/Hz}], \quad \forall l \in \psi.
\end{equation}

\subsubsection{Symbiotic Radio Network}

In SR networks, the SBDs operate passively and are unable to directly decode complex modulated signals from the environment. Instead, they embed their information onto the carrier of the incident signal which contains both private and common message components and reflect it toward their designated receiver, namely the SUE.

In addition to the backscattered signal from the desired \( \text{SBD}_i \), the SUE also receives a direct signal from the satellite, which occupies the same frequency band as the SBD's signal. Given the relatively higher power of the direct satellite link, it can be effectively eliminated using SIC. However, signals from other SBDs (i.e., \( j \neq i \)) introduce interference during the decoding of \( \text{SBD}_i \)'s signal at the SUE.

Accordingly, the SINR for decoding the \( i \)-th SBD’s signal is written as:
\begin{equation}
\label{eq:sinr_sbd_psd}
\gamma_{s,i}(t) =
\frac{\eta_i \left| h_i^r \mathbf{h}_i^H \mathbf{R}^{\mathrm{Out}}(t) \right|^2}
{\displaystyle \sum_{j \in \upsilon, j \neq i} \eta_j \left| h_j^r \mathbf{h}_j^H \mathbf{R}^{\mathrm{Out}}(t) \right|^2 + \textcolor{black}{N_0} },
\end{equation}
and the corresponding achievable backscatter data rate per unit bandwidth is expressed as \cite{8907447}:
\textcolor{black}{
\begin{equation}
\label{eq:R_s_psd}
R_{\mathrm{SR},i}(t) = \frac{\tau_i^{\mathrm{BD}}}{K} \log_2 \!\left( 1 + K\gamma_{s,i}(t) \right), \quad \forall i \in \upsilon.
\end{equation}
}
\textcolor{black}{The total backscatter sum-rate across all SBDs in the network is then \( R_{\mathrm{SR}}(t) = \sum_{i \in \upsilon} R_{\mathrm{SR},i}(t) \).}
\subsection{Energy Harvesting and Consumption Model}

\subsubsection{Energy Consumption}

To evaluate the total power requirements in the proposed Active SIM-assisted satellite communication system, it is essential to consider multiple energy-consuming components. These include the power loss in the satellite’s transmission circuits, the P.A consumption at the satellite, the active power required for amplification and reconfiguration in the ASIM, as well as the signal processing overhead associated with multi-layer beamforming.

Note that the power consumption of the user equipment is assumed to be negligible. Accordingly, the total power consumption is expressed as \cite{niu2023active}:

\begin{equation}
\label{eq:P_sum}
\begin{gathered}
  P_{\text{total}} = \vartheta_{\text{sat}}\| \mathbf{W} \|_F^2  + \vartheta_{\text{SIM}} \left( {\|\mathbf{T} \mathbf{F} \mathbf{W}\|_F^2 + \|\mathbf{T}\|_F^2 \sigma_{\mathrm{SIM}}^2} \right)\hfill \\ 
  \,\,\,\,\,\,\,\,\,\,\,\,\,\,\,\,+P_{c,\text{sat}} + P_{c,\text{SIM}},  \hfill \\ 
\end{gathered} 
 \end{equation}

where
\begin{align}
P_{c,\text{sat}} &= N P_D^{\text{sat}} + P_C^{\text{sat}}, \label{eq:P_sum2}\\
P_{c,\text{SIM}} &= M \left( P_{\text{phs}} + P_{\text{amp}} + P_{DC}^{\text{SIM}} + P_{\text{proc}} \right). \label{eq:P_c_sim}
\end{align}

In Eq.~\eqref{eq:P_sum}, the coefficients $\vartheta_{\text{sat}} = 1/\eta_{\text{PA}}^{\text{sat}}$ and $\vartheta_{\text{SIM}} = 1/\eta_{\text{PA}}^{\text{SIM}}$ represent the inverse of the power amplifier efficiencies at the satellite and the ASIM, respectively. They serve to quantify the excess input power resulting from non-ideal amplifier performance.  \textcolor{black}{These coefficients, \({\vartheta _{\mathrm{sat}}}\) and \({\vartheta _{\mathrm{SIM}}}\), are fixed, hardware-dependent efficiency parameters established by the power amplifier's design.}

The term $P_D^{\text{sat}}$ denotes the dissipation power per antenna element at the satellite, typically in the range of 50–100~mW. $P_C^{\text{sat}}$ corresponds to the power consumption of the satellite’s static circuits and control systems, generally between 2-5~W.

On the ASIM side, various hardware elements contribute to the total power consumption. The phase shifters consume a power of $P_{\text{phs}}$, typically between 1.5 and 7.8~mW per element, depending on the phase resolution (3–6 bits) \cite{huang2019reconfigurable}. The term $P_{\text{amp}}$ denotes the power required by integrated amplifiers for each active element, usually in the range of 5–15~mW to achieve a gain of 5–10~dB \cite{shi2025energy}. The DC biasing power, $P_{DC}^{\text{SIM}}$, is needed for biasing each tunable meta-atom and typically falls within 5–10~mW \cite{niu2023active}. Finally, $P_{\text{proc}}$ represents the computational power needed for impedance control and electromagnetic tuning at the element level, estimated to be around 1–3~mW per element \cite{niu2023active}.

\textcolor{black}{
To assess robustness, we note that moderate variations in hardware-related
parameters such as power amplifier efficiency and circuit power consumption
(e.g., within $\pm10$--$15\%$ of the nominal values reported in
\cite{huang2019reconfigurable,shi2025energy,niu2023active}) mainly result in
affine shifts of the total power consumption and do not alter the relative
energy efficiency trends or comparative performance conclusions.
}

\subsubsection{Energy Harvesting}

The harvested energy from solar radiation, which serves as the primary energy source for the satellite, is modeled by \cite{boddu2019solar}:
\begin{equation}
P_{\text{harvest}} = A_{\text{solar}} \cdot E_{\text{solar}} \cdot \eta_{\text{solar}} \cdot \text{PR} \cdot f_{\text{eclipse}},
\end{equation}
where $A_{\text{solar}} = x \times y$ is the total surface area of the solar cell array (in m$^2$), directly determining the collection capacity. The term $E_{\text{solar}}$ represents the solar irradiance under the Air Mass Zero (AM0) spectrum, which is approximately 1360~W/m\textsuperscript{2} in LEO conditions.

The solar cell efficiency $\eta_{\text{solar}}$ is typically 28\%–32\% for advanced multijunction cells. The performance ratio (PR), accounting for losses such as temperature, wiring, and DC–RF conversion, is about 75\%–85\% \cite{goh2015space}. The eclipse factor $f_{\text{eclipse}}$, representing the fraction of orbital time in sunlight, usually ranges from 0.6 to 0.8 for LEO satellites \cite{goh2015space}.

\section{Problem Formulation}

In the preceding sections, we developed a comprehensive model for the target satellite
communication system, aiming to deliver optimized service to legitimate ground users via
the RSMA protocol, while simultaneously supporting energy-limited IoT devices through a
CSR framework. The overall design seeks to jointly maximize the
sum-rate of the RSMA users and the backscatter throughput of the SBDs, while minimizing
total energy consumption, especially the transmit power of the satellite and active
reflection power of the ASIM.

\textcolor{black}{To enable a meaningful trade-off between energy efficiency and spectral efficiency while ensuring dimensional consistency, the objective function is formulated using normalized quantities. Specifically, the normalized total power consumption is defined as \(P^{dl}_{\mathrm{total}} = \frac{P_{\mathrm{total}}}{P^{\max}_{\mathrm{total}}}\), where \(P^{\max}_{\mathrm{total}}\) is a fixed reference value corresponding to the system power budget. Similarly, the achievable rates are normalized as \(R^{dl}_{\mathrm{sum}}(t) = \frac{R_{\mathrm{sum}}(t)}{R^{\max}_{\mathrm{sum}}}\) and \(R^{dl}_{\mathrm{SR}}(t) = \frac{R_{\mathrm{SR}}(t)}{R^{\max}_{\mathrm{SR}}}\), with the denominators being fixed system parameters representing the maximum achievable rates.}

The corresponding optimization problem is as follows:
\begin{subequations}
\label{eq:CVX}
\begin{align}
&\min_{\substack{ \mathbf{\Phi}^{(q)}, \mathbf{W}, \tau_i^{\mathrm{BD}},\\ \tau_i^{\mathrm{EH}}, 
C_l(t), \sigma_c, \sigma_l, \eta_i}} 
  P^{dl}_{\mathrm{total}} -  \left( R^{dl}_{\mathrm{sum}}(t) + R^{dl}_{\mathrm{SR}}(t) \right)
\\
&\text{s.t.} \notag \\[2mm]
&\quad P_{\mathrm{total}} \leq P_{\mathrm{harvest}}, \label{eq:subeq17b} \\[1mm]
&\quad \sigma_c + \sum_{l=1}^L \sigma_l = 1, 
\qquad \sigma_c\ge 0,\;\sigma_l\ge 0, \label{eq:subeq17c} \\[1mm]
&\quad \| \mathbf{W} \|_F^2 \leq  P_{\max}^{\mathrm{sat}}, \label{eq:subeq17d} \\
&\quad 
\|\mathbf{T} \mathbf{F} \mathbf{W}\|_F^2 + \|\mathbf{T}\|_F^2 \sigma_{\mathrm{SIM}}^2 \le P_{\max}^{\mathrm{SIM}}, 
\label{eq:subeq17e}
\\[1mm]
&\quad 
\sum_{l=1}^L C_l(t) 
\leq 
\min_{l\in\psi} \big\{ R_{c,l}(t) \big\},
\quad \forall t,
\label{eq:subeq17f}
\\[1mm]
&\quad 
R_{\mathrm{sum}}^{\mathrm{th}}(t)
\le R_{\mathrm{sum}}(t)
\le \sum_{l=1}^L \big( C_l(t) + R_{p,l}(t) \big),
\quad \forall t,
\label{eq:subeq17g}
\\[1mm]
&\quad
R_{\mathrm{SR}}^{\mathrm{th}}(t) 
\le 
R_{\mathrm{SR},i}(t) 
\le 
\frac{ \tau_i^{\mathrm{BD}}}{K}
\log_2 \big( 1 + K\gamma_{s,i}(t) \big),
\quad \forall i,t,
\label{eq:subeq17h}
\\[1mm]
&\quad 
0 \le \angle\Phi_m^{(q)} \le 2\pi, 
\qquad \forall m,q, 
\label{eq:subeqSIMi}
\\[1mm]
&\quad 
0 \le \big|\Phi_m^{(q)}\big| 
\le \sqrt{P_{\max}^{\mathrm{SIM}}}, 
\qquad \forall m,q, 
\label{eq:subeqSIMj}
\\[1mm]
&\quad 
0 \le \eta_i \le 1, 
\qquad i\in\upsilon, 
\label{eq:subeq17k}
\\[1mm]
&\quad 
\varepsilon_{\mathrm{SBD}_i} 
\le 
\Gamma \tau_i^{\mathrm{EH}}
\left| \mathbf{h}_i^H \mathbf{R}^{\mathrm{Out}}(t) \right|^2,
\qquad i\in\upsilon,
\label{eq:subeq17l}
\\[1mm]
&\quad 
\tau_i^{\mathrm{EH}} + \tau_i^{\mathrm{BD}} = 1,
\qquad i\in\upsilon.
\label{eq:subeq17x} 
\end{align}
\end{subequations}
where constraint~\eqref{eq:subeq17b} ensures that the total power consumed by the satellite and the active SIM does not exceed the harvested solar energy \(P_{\mathrm{harvest}}\). {Signal power allocation for RSMA is enforced through. 
Constraint~\eqref{eq:subeq17c} normalizes the power-splitting coefficients 
\(\sigma_c\) and \(\{\sigma_l\}_{l=1}^L\). 
Meanwhile, Constraints~\eqref{eq:subeq17d} and \eqref{eq:subeq17e} impose transmit power limits on the satellite precoder \(\mathbf{W}\) and the active 
SIM layers \(\mathbf{\Phi}^{(q)}\) by constraining both the amplified forward signal 
\(\mathbf{\Phi}^{(q)} \mathbf{F}\mathbf{W}\) and the SIM-induced distortion power, respectively.
With \eqref{eq:subeq17e} accounting for both the amplified signal power and noise introduced by active metasurface elements. Constraints~\eqref{eq:subeq17f}–\eqref{eq:subeq17h} guarantee reliable decodability and minimum QoS for RSMA and commensal SR users. Constraints~\eqref{eq:subeqSIMi} and~\eqref{eq:subeqSIMj} restrict the feasible phase \(\angle \Phi_m^{(q)}\) and amplitude \(|\Phi_m^{(q)}|\) ranges of each meta-atom \(m\) in each SIM layer \(q\), reflecting hardware limitations. Constraints~\eqref{eq:subeq17k}--\eqref{eq:subeq17x} govern the corresponding energy efficiency \(\eta_i\), the energy harvesting of \(\text{SBD}_i\), and their backscatter communication and energy harvesting time allocation.

Due to the non-convex and highly complex nature of the optimization problem \ref{eq:CVX}, we adopt a mathematical convexification technique called BCD-SCA to iteratively approximate the problem with convex subproblems. To effectively solve these approximations in a model-free manner, we employ two advanced deep reinforcement learning algorithms: MCPPO and MA-CSAC.

\section{Convexification via BCD-SCA}

The BSD-SCA methodology tackles the non-convex optimization problem in~\eqref{eq:CVX} by iteratively approximating non-convex components with convex surrogates and solving tractable subproblems to ensure efficient convergence.

The main source of non-convexity lies in the objective function, which combines normalized power consumption and sum-rate terms, where both \(R_{\mathrm{SUM}}\) and \(R_{\mathrm{SR}}\) involve \(\log_{2}(1+\text{SINR})\) expressions with SINR ratios of coupled variables.

Constraint \eqref{eq:subeq17f} represents the pointwise minimum of rate functions across multiple users, which is inherently non-convex due to the minimum operation. Moreover, constraint \eqref{eq:subeq17g} involves the sum of non-concave logarithmic functions, further contributing to non-convexity. Similarly, constraint \eqref{eq:subeq17h} is non-convex because it contains a fractional SINR expression inside the logarithmic function, introducing coupled variables in both the numerator and the denominator.

The SINR expressions \( \gamma_{c,l}(t), \gamma_{p,l}(t), \gamma_{s,i}(t) \) are inherently non-convex, as they are ratios of functions involving interdependent variables. The combination of these ratios with logarithmic functions amplifies their non-convex nature.

The amplitude and phase constraints on the ASIM entries, \( |\Phi_m^{(q)}| \) and \( \angle\Phi_m^{(q)} \), when combined with signal propagation effects, result in non-convexity. The joint consideration of amplitude and phase changes leads to a complex non-convex optimization landscape.

To overcome the challenges posed by the non-convex components, we partition the optimization variables into three disjoint blocks and adopt a BCD framework. Within each block, we apply SCA to approximate the corresponding non-convex terms. The SCA method allows us to transform the non-convex components into convex surrogates, enabling us to solve the optimization problem iteratively.

\begin{itemize}
  \item \textbf{Block 1:} Satellite precoder and power variables ($\mathbf{W}, \sigma_c, \sigma_l$)
  \item \textbf{Block 2:} ASIM tuning matrices ($\mathbf{\Phi}^{(q)}$)
  \item \textbf{Block 3:} Backscatter parameters ($\tau_i^{\mathrm{EH}}, \tau_i^{\mathrm{BD}}, \eta_i$)
\end{itemize}

\textcolor{black}{
Each block is optimized alternately while the other blocks are kept fixed, and
the non-convex components are convexified via SCA such that, under the imposed
domain constraints (e.g., nonnegativity of signal powers, positivity of
interference terms, bounded SIM amplitudes, and normalized power coefficients),
each surrogate constitutes a global lower bound that is tight at the current
iterate, thereby ensuring that every block-wise subproblem in the BCD framework
is a well-defined convex program.
}

\subsection*{\textbf{Block 1}: Satellite Precoder and Power Allocation}

Optimized Variables: $\mathbf{W}, \sigma_c, \sigma_l$ \\ 
{Fixed Variables:} $\mathbf{\Phi}^{(q)}, \tau_i^{\mathrm{EH}}, \tau_i^{\mathrm{BD}}, \eta_i$

In Block 1, SINR expressions are approximated with quadratic transformations using auxiliary variables \( y_{c,l} \) and \( y_{p,l} \). Achievable rates are linearized via first-order Taylor expansion. The pointwise minimum in the common rate constraint is reformulated with a threshold \( \bar{R}_c \).

\paragraph{Quadratic Transformation of SINR}

\textcolor{black}{For each SINR term $\gamma_{c,l}=A_l/B_l$, where $A_l\ge 0$ denotes the desired signal power and $B_l>0$ represents the interference-plus-noise term, we apply a quadratic SCA-based transformation by introducing an auxiliary variable $y_{c,l}$. This yields the lower bound $\gamma_{c,l} \ge 2y_{c,l}\sqrt{A_l}-y_{c,l}^2 B_l$, with the auxiliary variable updated at each SCA iteration as $y_{c,l}^{(k)}=\sqrt{A_l^{(k-1)}}/B_l^{(k-1)}$, which guarantees tightness at the current iterate and provides a global lower bound.} \textcolor{black}{For fixed $y_{c,l}$, the expression $2y_{c,l}\sqrt{A_l}-y_{c,l}^2 B_l$ is jointly concave in the optimization variables (with $\sqrt{A_l}$ further linearized via a first-order Taylor expansion), thereby rendering the resulting subproblem convex.}

\paragraph{First-Order Rate Approximation}
We linearize the rate function \( R_{c,l}(\gamma_{c,l}) \) around the current SINR value \( \gamma_{c,l}^{(k)} \) using a first-order Taylor expansion:
\begin{equation}
R_{c,l} \approx R_{c,l}^{(k)} + \frac{1}{(1+\gamma_{c,l}^{(k)})\ln 2} \bigl(\gamma_{c,l} - \gamma_{c,l}^{(k)}\bigr).
\end{equation}
This approximation simplifies the rate function to a linear form, making it more tractable for optimization.

\paragraph{Min-Rate Reformulation}  
To simplify the non-convex pointwise minimum in the constraint \( \min_{l} R_{c,l}(t) \), we introduce a common rate threshold \( \bar{R}_c \) and reformulate the constraint as \( R_{c,l} \ge \bar{R}_c, \ \forall l \). This transforms the original minimum operation into a set of linear constraints, making the problem more tractable.

\paragraph{Resulting Subproblem 1}
After these transformations, we obtain a convex subproblem:
\begin{equation}
\min_{\mathbf{W},\, \sigma_c,\, \sigma_l,\, \bar{R}_c} \; P^{dl}_{\mathrm{total}} - \bigl( \bar{R}^{dl}_c + \sum_{l} R^{dl}_{p,l} \bigr)
\end{equation}
subject to linearized SINR and rate constraints for both common and private messages.

\subsection*{\textbf{Block 2}: Active SIM Tuning}

{Optimized Variables:} $\mathbf{\Phi}^{(q)} = \mathrm{diag}(\Phi_1^{(q)}, \ldots, \Phi_M^{(q)})$\\ 
{Fixed Variables:} $\mathbf{W}, \sigma_c, \sigma_l, \tau_i^{\mathrm{EH}}, \tau_i^{\mathrm{BD}}, \eta_i$

In Block 2, we decouple amplitude and phase components by expressing each complex entry in Cartesian form. The quadratic constraint becomes convex.

\paragraph{Majorization-Minimization (MM) Linearization}
Expressions such as $|\mathbf{g}_l^H \mathbf{T} \mathbf{F} \mathbf{w}_c|^2$,
are approximated using a linear surrogate at iteration \( \phi^{(k)} \). This allows us to replace the non-linear terms with their linear approximations, making the optimization tractable.

\paragraph{Amplitude-Phase Decoupling}
We use variable substitution to decouple the amplitude and phase:
\begin{equation}
|\Phi_m^{(q)}| \cos(\angle\Phi_m^{(q)}) = a_m^{(q)}, \quad |\Phi_m^{(q)}| \sin(\angle\Phi_m^{(q)}) = b_m^{(q)},
\end{equation}
with the convex quadratic constraint $
(a_m^{(q)})^2 + (b_m^{(q)})^2 \le P_{\max}^{\mathrm{SIM}}.
$
This decoupling transforms the non-convex amplitude-phase constraints into convex constraints.

\paragraph{Resulting Subproblem 2}
We now have a convex program over \( (a_m^{(q)}, b_m^{(q)}) \) to minimize the linearized objective subject to power and SINR constraints.

\subsection*{\textbf{Block 3}: Backscatter Parameters}

{Optimized Variables:} $\tau_i^{\mathrm{EH}}, \tau_i^{\mathrm{BD}}, \eta_i$ \\ 
{Fixed Variables:} $\mathbf{W}, \sigma_c, \sigma_l, \mathbf{\Phi}^{(q)}$

In Block 3, we adopt the \textit{Augmented Lagrangian} (AL) method to address the constrained optimization associated with backscatter parameters. This method efficiently incorporates both the EH and SR constraints using penalty functions and Lagrange multipliers.

\paragraph{AL Formulation}
The AL function is defined as:
\begin{equation}
\mathcal{L}_\rho(\tau, \eta, \lambda) = f(\tau, \eta) + \frac{\rho}{2} \sum_{i} \left( h_i(\tau_i^{\mathrm{EH}}, \tau_i^{\mathrm{BD}}) + \frac{\lambda_i}{\rho} \right)^2,
\end{equation}
where \( f(\tau, \eta) \) denotes the convexified objective function, \( h_i(\cdot) = \tau_i^{\mathrm{EH}} + \tau_i^{\mathrm{BD}} - 1 \) represents the equality constraint, \( \lambda_i \) are the Lagrange multipliers, and \( \rho > 0 \) is the penalty parameter.

\paragraph{Constraint Handling}
The EH and SR constraints are incorporated as convex approximations in the AL framework. The solution proceeds iteratively via:
\begin{equation}
(\tau^{(k+1)}, \eta^{(k+1)}) = \arg\min_{\tau, \eta} \mathcal{L}_{\rho^{(k)}}(\tau, \eta, \lambda^{(k)}),
\end{equation}
with Lagrange multipliers updated as:
\begin{equation}
\lambda_i^{(k+1)} = \lambda_i^{(k)} + \rho^{(k)} \cdot h_i\left( \tau_i^{\mathrm{EH},(k+1)}, \tau_i^{\mathrm{BD},(k+1)} \right),
\end{equation}
and penalty parameter adjusted via $
\rho^{(k+1)} = \alpha \cdot \rho^{(k)}, \quad \text{with } \alpha > 1$.

\paragraph{Convex Subproblem}
The inner optimization is convex and can be efficiently solved using:
\begin{itemize}
  \item First-order Taylor expansions for convexifying EH constraints;
  \item Geometric programming techniques for SR rate terms;
  \item Quadratic penalties to enforce time-sharing constraints.
\end{itemize}

\paragraph{Resulting Subproblem 3}
The AL method transforms the original constrained problem into a sequence of unconstrained convex subproblems:
\begin{equation}
\min_{\tau, \eta} \mathcal{L}_\rho(\tau, \eta, \lambda),
\end{equation}
with adaptive updates of multipliers and penalty parameters ensuring constraint satisfaction and convergence.

Based on the algorithm in Algorithm \ref{tab:BCD-RCD}, the problem is transformed into a convex optimization problem, which can be simulated in MATLAB using CVX.

\begin{algorithm}{\small
\caption{BCD-SCA Algorithm}
\label{tab:BCD-RCD}
\begin{algorithmic}[1]  
  \STATE \textbf{Initialize:} $\mathbf{W}^{(0)}, \mathbf{\Phi}^{(0)}, \tau^{(0)}, \eta^{(0)}, k=0$
  \STATE \textbf{Set convergence thresholds:} \( \epsilon, \delta \)
  \REPEAT
    \STATE \textbf{[Block 1] Precoder and Power Allocation}
    \STATE \quad Update \( y_{c,l}, y_{p,l} \) using quadratic transformations
    \STATE \quad Solve Subproblem 1

    \STATE \textbf{[Block 2] ASIM Tuning}
    \STATE \quad Linearize SINRs at \( \mathbf{\Phi}^{(k)} \) using MM and decouple amplitude-phase components
    \STATE \quad Solve Subproblem 2

    \STATE \textbf{[Block 3] Backscatter}
    \STATE \quad Initialize AL parameters: \( \lambda^{(0)} = 0, \rho^{(0)} > 0, j=0 \)
    \REPEAT
      \STATE \quad Solve \(\min_{\tau, \eta} \mathcal{L}_{\rho^{(j)}}(\tau, \eta, \lambda^{(j)})\) subject to constraints
      \STATE \quad Update \( \lambda_i^{(j+1)} = \lambda_i^{(j)} + \rho^{(j)} h_i(\tau_i^{\mathrm{EH}}, \tau_i^{\mathrm{BD}}) \)
      \STATE \quad Increase penalty: \( \rho^{(j+1)} = \alpha \rho^{(j)} \) with \(\alpha > 1\)
      \STATE \quad \( j \gets j + 1 \)
    \UNTIL{\( \| h_i(\tau_i^{\mathrm{EH}}, \tau_i^{\mathrm{BD}}) \|_2 < \epsilon_{\text{AL}} \)}

    \STATE  \( k \gets k + 1 \)
  \UNTIL{\( |\text{Obj}^{(k)} - \text{Obj}^{(k-1)}| < \epsilon \) \textbf{,} \( \| \mathbf{W}^{(k)} - \mathbf{W}^{(k-1)} \|_F < \delta \)}
\end{algorithmic}
}\end{algorithm}

\section{Deep Reinforcement Learning Methods}
\label{sec:DRL_framework}

To tackle the constrained non-convex optimization in \eqref{eq:CVX}, we reformulate it as a \textit{constrained multi-agent Markov Decision Process (CMDP)}, capturing the coupling among beamforming, ASIM phase shifts, and time allocation. This enables policy-gradient learning while preserving the problem structure.

\textcolor{black}{
We employ two complementary DRL engines, MA--CSAC and MCPPO, because the control task is intrinsically multi-modal and multi-timescale: RSMA precoding and per-element SIM tuning require low-variance, sample-efficient continuous control, whereas backscatter scheduling, energy-neutrality, and long-horizon QoS impose strict CMDP constraints with slower discrete dynamics. 
}\textcolor{black}{
In practice, the two algorithms exhibit distinct and complementary convergence behaviors. MCPPO (on-policy) provides {fast initial convergence} due to its clipped surrogate structure and disciplined on-policy updates, although its iterates show higher variance in early episodes. MA--CSAC (off-policy, CTDE), equipped with heterogeneous action priors (Gaussian for beamforming, von Mises for phases, Beta for gains), a centralized twin-Q critic with Lagrangian penalties, and entropy-regularized updates with adaptive multipliers, achieves {smoother and more stable long-term learning}, ultimately reaching higher final rewards in our experiments.
}\textcolor{black}{
MCPPO complements this by ensuring constraint-aware policy improvement over long horizons. Our MCPPO design integrates a multi-constraint clipped surrogate objective, discounted constraint costs, adaptive Lagrange-multiplier updates tuned for on-policy sampling, and modified advantage estimation/clipping to reconcile fast satellite symbol timescales with slower backscatter slots.
}\textcolor{black}{
Together, these design choices yield a practical hybrid solution: {rapid early progress via MCPPO, high-quality asymptotic refinement via MA--CSAC, and consistent satisfaction of CMDP constraints}, forming a robust learning framework for the LEO–ASIM–RSMA control problem.
}

\begin{table}[H]
\centering
\caption{Key Characteristics of MA-CSAC and MCPPO}
\label{tab:comparison}
\begin{tabular}{|c|c|c|c|}
\hline
\textbf{Attribute} & \textbf{MA-CSAC} & \textbf{MCPPO} \\ \hline
Learning paradigm & Off-policy & On-policy \\ \hline
Exploration & Entropy-based & Stochastic sampling \\ \hline
Constraint handling & Distributed critics & Centralized penalty \\ \hline
Sample efficiency & High & Moderate \\ \hline
Convergence & Fast initially & Strong asymptotically \\ \hline
Stability & Moderate & High \\ \hline
Action space & Continuous & Continuous/Discrete \\ \hline
Parallelization & High & Moderate \\ \hline
\end{tabular}
\end{table}

\subsection{Constrained Markov Decision Process (CMDP)}
\label{subsec:CMDP}

The CMDP is defined by the tuple $(\mathcal{S}, \mathcal{A}, \mathcal{P}, r, \{\mathcal{C}_j,\bar{c}_j\}_{j=1}^J, \gamma)$, where $\mathcal{S}$ is the state space, $\mathcal{A}$ the joint action space, $\mathcal{P}$ the transition probability, $r$ the reward, and each constraint $\mathcal{C}_j$ has a threshold $\bar{c}_j$. At each step $t$, the environment yields $(\mathbf{s}_t,\mathbf{a}_t,r_t,\mathbf{s}_{t+1},\{c_t^{(j)}\}_{j=1}^J)$ with $c_t^{(j)}=\mathcal{C}_j(\mathbf{s}_t,\mathbf{a}_t)$.

The state $\mathbf{s}_t$ includes time-varying channel parameters $\{\mathbf{g}_l(t)\}$, beamforming matrix $\mathbf{F}(t)$, SIM responses $\{\mathbf{H}^{(q)}(t)\}$, user channels $\{\mathbf{h}_i(t)\}$, and time allocations $\{\tau_i(t)\}$. The joint action $\mathbf{a}_t$ controls transmission, SIM reconfiguration, and CSR resources:\textcolor{black}{
\begin{equation} \label{eq:action_cmdp}
\mathbf{a}_t =
\{ \mathbf{W}_t, \{\mathbf{\Phi}^{(q)}_t\}, \boldsymbol{\tau}_t,
\boldsymbol{\eta}_t, \boldsymbol{\sigma}_t, \mathbf{C}_t \}.
\end{equation}
}
\textcolor{black}{
Here, the time index $t$ denotes the decision epoch at which the optimization
variables are updated based on the observed system state and does not imply
time-varying satellite geometry or deterministic channel knowledge assumptions,
while the action space is strictly limited to physically controllable variables
that can be adjusted online by the satellite and ASIM controller, with channel
realizations, hardware efficiencies, and propagation parameters treated as
exogenous state variables rather than optimization actions.
}

The reward balances SE and power efficiency:
\begin{equation} \label{eq:reward_cmdp}
r_t = \big( R^{dl}_{\mathrm{SUM}}(t) + R^{dl}_{\mathrm{SR}}(t) \big) - P^{dl}_{\mathrm{total}} - \sum_{j=1}^J \lambda_j \max(0,\mathcal{V}_t^{(j)}),
\end{equation}
where $\mathcal{V}_t^{(j)} = c_t^{(j)}-\bar{c}_j$. Lagrange multipliers $\lambda_j$ are updated via projected gradient ascent:
\begin{equation} \label{eq:lagrange_update}
\lambda_j \leftarrow \max\!\big(0,\lambda_j+\eta_\lambda \cdot \max(0,\mathcal{V}_t^{(j)}-\epsilon_{\mathrm{tol}})\big),
\end{equation}
with $\eta_\lambda>0$ the step size and $\epsilon_{\mathrm{tol}}$ a tolerance margin for minor violations, ensuring feasibility and stable learning.

\subsection{Multi-Agent Constraint Soft Actor-Critic (MA-CSAC)}
\label{subsec:MA_CSAC}

The proposed MA-CSAC extends constraint-aware multi-agent reinforcement learning \cite{shang2023constraint} by adapting SAC to collaborative settings with explicit constraints. Following the centralized training and decentralized execution (CTDE) framework, it uses decentralized stochastic actors for independent execution and a centralized critic that evaluates joint state-action pairs to improve sample efficiency and stability. Tailored for satellite-ASIM-backscatter systems, MA-CSAC handles heterogeneous actions (Gaussian for beamforming, von Mises for phase shifts, Beta for gain control) and enforces operational constraints via Lagrangian relaxation, ensuring scalable exploration with strict compliance.

\textcolor{black}{Each agent is assigned to a specific RSMA user along with its corresponding ASIM elements. This agent division reduces the dimensionality of individual state-action spaces, facilitating scalable learning. During offline training, the centralized critic aggregates joint state-action information, providing stable updates, while each actor executes independently during online inference, significantly minimizing communication overhead.}

Each agent maintains a stochastic policy $\pi_{\phi_a}$ that maximizes long-term expected return with entropy regularization, penalizing constraint violations:
{\small\begin{equation} \label{eq:macsac_obj_final}
\max_{\pi_a} \mathbb{E} \Big[ \sum_{t=0}^{\infty} \gamma^t \big( r_t + \alpha \mathcal{H}(\pi_a(\cdot|\mathbf{s}_t)) - \sum_{j=1}^J \lambda_j(t) c_t^{(j)} \big) \Big],
\end{equation}}
with entropy $\mathcal{H}(\pi_a)$ encouraging exploration, $\lambda_j(t)\ge0$ as Lagrange multipliers, and $c_t^{(j)}$ the instantaneous constraint costs.

Policy distributions match action types: Gaussian for transmitter beamforming/power, von Mises for ASIM phase shifts, and Beta for ASIM gain control (constraining gains to $[0,1]$).

\textcolor{black}{To address the credit assignment problem, each agent's advantage is computed based on its contribution to the global reward. This approach ensures that updates reflect the individual impact of agents on the overall system performance, promoting stable and efficient learning.}

The centralized critic uses a twin-Q network to estimate soft state-action values:
\begin{equation}
Q_{\zeta}(\mathbf{s}, \mathbf{a}) = \min_{k=1,2} f_{\zeta_k}\left(\mathbf{s}, \mathbf{a}^{\text{TX}}, \mathbf{a}^{\text{SIM-phase}}, \mathbf{a}^{\text{SIM-gain}}\right),
\end{equation}
with $\mathbf{a} = \{\mathbf{a}^{\text{TX}}, \mathbf{a}^{\text{SIM-phase}}, \mathbf{a}^{\text{SIM-gain}}\}$. The twin-Q structure reduces overestimation bias and improves stability.
\begin{algorithm}{\small
\caption{MA-CSAC with Lagrangian Constraint Handling}
\label{alg:MACSAC}
\begin{algorithmic}[1]
  \STATE \textbf{Initialize:} Actor parameters $\{\phi_i\}$, critic parameters $\{\zeta_k\}$, Lagrange multipliers $\{\lambda_j^{(0)}\}$, target Q-networks, and replay buffer $\mathcal{D}$.
  \FOR{episode $= 1$ \TO $E_{\max}$}
    \STATE Observe initial state $\mathbf{s}_0$
    \FOR{$t=0$ \TO $T-1$}
      \STATE Sample actions $\mathbf{a}_i \sim \pi_{\phi_i}(\cdot|\mathbf{s}_t)$ for $i \in \{\text{TX}, \text{ASIM-phase}, \text{SIM-gain}\}$
      \STATE Execute joint action $\mathbf{a}_t = \{\mathbf{a}_\text{TX}, \mathbf{a}_\text{SIM-phase}, \mathbf{a}_\text{SIM-gain}\}$, observe $r_t$, $\mathbf{s}_{t+1}$, and $\{c_t^{(j)}\}$
      \STATE Store transition $(\mathbf{s}_t, \mathbf{a}_t, r_t, \mathbf{s}_{t+1}, \{c_t^{(j)}\})$ in $\mathcal{D}$
    \ENDFOR
    \FOR{gradient step $= 1$ \TO $G$}
      \STATE Sample mini-batch $B \sim \mathcal{D}$
      \STATE Update critics: $\zeta_k \leftarrow \zeta_k - \eta_Q \nabla_{\zeta_k} J_Q(\zeta_k)$ for $k=1,2$
      \STATE Update actors: $\phi_i \leftarrow \phi_i + \eta_\phi \nabla_{\phi_i} J_\pi(\phi_i)$ for each agent $i$
      \STATE Soft-update targets: $\zeta_k^{\text{target}} \leftarrow \tau \zeta_k + (1 - \tau)\zeta_k^{\text{target}}$
      \STATE Update multipliers:
        \(
        \lambda_j \leftarrow \left[\lambda_j + \eta_\lambda \left(\bar{c}_j^{(B)} - \epsilon_{\mathrm{tol}}\right)\right]_+,
        \)
        where $\bar{c}_j^{(B)}$ is the mini-batch average of constraint $j$
    \ENDFOR
  \ENDFOR
\end{algorithmic}
}\end{algorithm}

\vspace{1mm}
Empirical results confirm that the combination of CTDE structure, entropy regularization, and individual advantage-based credit assignment accelerates convergence while maintaining strict compliance with constraints, enabling practical online deployment in real-time satellite operations.

\subsection{Multi-Constraint Proximal Policy Optimization (MCPPO)}
\label{subsec:MCPPO}

The MCPPO algorithm extends the standard PPO~\cite{schulman2017proximal} to explicitly handle multiple system-level constraints via Lagrangian-based penalty terms~\cite{achiam2017constrained}. In our satellite--SIM--SR setup, these constraints include transmit power limits, phase stability requirements, and amplification gain bounds, which are essential for maintaining reliable operation in heterogeneous action spaces. While the standard PPO formulation is insufficient for enforcing such constraints, MCPPO preserves PPO’s stability advantages, such as the clipped surrogate objective, while augmenting it with constraint penalties that guide the policy towards feasible actions. This hybrid design ensures that, even under the high-dimensional and non-convex nature of our problem, the learning process remains stable, constraint-compliant, and capable of delivering near-optimal performance. The resulting clipped objective with constraint penalties is formulated as:
\begin{multline}{\small \label{eq:mcppo_obj_final}
J^{\text{MCPPO}}(\theta) = \mathbb{E}_t \bigg[ \min\big( r_t(\theta) \hat{A}_t, \text{clip}(r_t(\theta), 1-\epsilon, 1+\epsilon) \hat{A}_t \big) \\
- \sum_{j=1}^{J} \lambda_j \max\big(0, \hat{C}_t^{(j)} - \bar{c}_j \big) \bigg]
}\end{multline}

where $r_t(\theta) = \pi_\theta(\mathbf{a}_t|\mathbf{s}_t) / \pi_{\theta_{\text{old}}}(\mathbf{a}_t|\mathbf{s}_t)$ is the probability ratio, $\hat{A}_t$ denotes the Generalized Advantage Estimator (GAE), $\hat{C}_t^{(j)} = \sum_{k=t}^{\infty} \gamma^{k-t} c_k^{(j)}$ represents the discounted cumulative cost of the $j$-th constraint, $\bar{c}_j$ is the threshold for the $j$-th constraint, and $\lambda_j \ge 0$ is the associated Lagrange multiplier, adjusted adaptively to enforce feasibility. $\epsilon_{\mathrm{tol}}$ is a small tolerance used to ignore negligible constraint violations and improve numerical stability.

\begin{algorithm}{\small
\caption{MCPPO with Adaptive Multipliers}
\label{alg:MCPPO}
\begin{algorithmic}[1]
  \STATE \textbf{Initialize:} Policy $\pi_{\theta}$, value network $V_{\psi}$, multipliers $\bm{\lambda}^{(0)}$
  \FOR{iteration = 1 \textbf{to} $I_{\max}$}
    \STATE Collect trajectories $\{\tau_i\}$ using $\pi_{\theta_{\text{old}}}$
    \STATE Compute advantages $\hat{A}_t$ and constraint costs $\hat{C}_t^{(j)}$
    \FOR{epoch = 1 \textbf{to} $K$}
      \STATE Update policy: $\theta \leftarrow \theta + \eta_\theta \nabla_\theta J^{\text{MCPPO}}(\theta)$
      \STATE Update value network: $\psi \leftarrow \psi - \eta_\psi \nabla_\psi \|V_\psi(\mathbf{s}_t) - \hat{V}_t\|^2$
    \ENDFOR
    \STATE Update 
    \(
      \lambda_j \leftarrow \left[ \lambda_j + \eta_\lambda \left( \mathbb{E}_t[\hat{C}_t^{(j)}] - \bar{c}_j - \epsilon_{\mathrm{tol}} \right) \right]_+
    \)
    \COMMENT{Projected to ensure $\lambda_j \ge 0$}
  \ENDFOR
\end{algorithmic}
}\end{algorithm}

{\color{black}
\subsection{Theoretical Remarks and Asymptotic Properties}
\label{subsec:theory}
This subsection provides theoretical remarks that contextualize the empirical
behavior of the proposed MCPPO and MA-CSAC algorithms.
Due to the inherently nonconvex nature of the considered CMDP, the use of
function approximation, and the presence of coupled multi-agent dynamics,
strong convergence guarantees or global optimality results are generally
intractable.
Instead, we summarize standard sufficient conditions under which
policy-gradient-based primal--dual reinforcement learning methods exhibit
non-degrading expected performance and asymptotic constraint satisfaction in
expectation.

The following assumptions are commonly adopted in the analysis of constrained
policy optimization and actor--critic methods, and apply to both MCPPO and
MA-CSAC.

\textbf{Assumption 1 } (Lipschitz Regularity).
The reward function $r(\mathbf{s},\mathbf{a})$ and each constraint cost
$\mathcal{C}_j(\mathbf{s},\mathbf{a})$ are Lipschitz continuous with respect to
the state--action pair, i.e.,
\begin{equation}
\|r(\mathbf{s},\mathbf{a}) - r(\mathbf{s}',\mathbf{a}')\|
\leq L_r \big(\|\mathbf{s}-\mathbf{s}'\| + \|\mathbf{a}-\mathbf{a}'\|\big),
\end{equation}
and similarly for $\mathcal{C}_j$ with constants $L_j$.

\textbf{Assumption 2 } (Bounded Variance and Approximation Error).
The stochastic advantage estimators for both the reward and constraint costs
have bounded variance, and the error induced by function approximation is
uniformly bounded, i.e.,
\begin{equation}
\mathbb{V}\mathrm{ar}[\hat{A}_t] \leq \sigma_A^2, \quad
\mathbb{V}\mathrm{ar}[\hat{C}_t^{(j)}] \leq \sigma_C^2,
\end{equation}
with a finite approximation error $\epsilon_{\mathrm{approx}}$.

Under these assumptions, standard results from trust-region-based and
primal--dual policy optimization yield the following properties.

\textbf{Proposition 1 } (Expected Performance Non-Degradation).
For sufficiently small policy update steps, the expected Lagrangian objective
$\mathcal{L}(\theta)$ associated with the CMDP satisfies
\begin{equation}
\mathbb{E}[\mathcal{L}(\theta_{k+1})]
\geq
\mathbb{E}[\mathcal{L}(\theta_k)]
-
\mathcal{O}(\epsilon_{\mathrm{approx}}),
\end{equation}
meaning that the expected performance does not degrade across iterations,
up to approximation errors.
This property follows from first-order policy improvement bounds and
trust-region arguments commonly used in PPO- and SAC-based methods.

\textbf{Proposition 2 } (Asymptotic Constraint Satisfaction in Expectation).
With properly chosen learning rates for the primal policy and dual variables,
the time-averaged expected constraint costs satisfy
\begin{equation}
\limsup_{T \to \infty}
\frac{1}{T}
\sum_{t=1}^T
\mathbb{E}[c_t^{(j)}]
\leq
\bar{c}_j + \epsilon_{\mathrm{tol}}, \quad \forall j,
\end{equation}
where $\epsilon_{\mathrm{tol}} > 0$ captures the residual error induced by
stochasticity and function approximation.
}

\section{Simulation Results}

This section presents the simulation setup used to evaluate the proposed MA-CSAC, MCPPO, and BCD-SCA algorithms. The scenario considers a LEO satellite at 500 km altitude integrated with ASIM, serving multiple randomly distributed ground users and SBDs.  

The system operates in the Ka-band at 20 GHz, with Rician fading channels to capture LoS and multipath effects. Signal attenuation includes path loss, frequency-dependent fading, thermal noise, and atmospheric absorption. RSMA enables simultaneous transmission of common and private messages, supported by ASIM beam steering and SBD backscatter modulation.  

BCD-SCA is implemented in MATLAB R2024a with CVX/SDPT3 to iteratively optimize beamforming vectors, SIM phase shifts, and backscatter coefficients under system constraints. MCPPO and MA-CSAC are DRL methods implemented in Python 3.10 with TensorFlow 2.13 and GPU acceleration (NVIDIA RTX 4050); all random generators (Python, NumPy, TensorFlow) were seeded with values 1 to 15 across 15 independent runs, as listed in Table~\ref{tab:training}. Parallel episodes accelerate convergence while jointly optimizing transmission parameters for SE and EE. Key simulation parameters are summarized in Table~\ref{tab:sim_parameters}, and DRL training hyperparameters are listed in Table~\ref{tab:training}. 

\begin{table}
    \centering
    \caption{Simulation Parameters.}
    \begin{tabular}{|c|c|c|c|}
        \Xhline{1.2pt} 
        \textbf{Parameter} & \textbf{Value} & \textbf{Parameter} & \textbf{Value}\\
        \Xhline{1.2pt} 
        x & 0.5 m &  pathloss exp. & 2.2 \\
        \hline
        y & 1 m &  $E_{\text{solar}}$ & 1360 W/m\textsuperscript{2} \\
        \hline
        $\Gamma$ &  0.8 & $\eta_{\text{solar}}$  & 32\%  \\
        \hline
	  $f_{\text{eclipse}}$ & 0.7 & PR & 80 \% \\
        \hline
        N & 32  &  M  & 128  \\
        \hline
        Freq. & 20 GHz & $B$ & 10 MHz \\
        \hline
        Q & 4 & Sat. altitude & 500 Km \\
        \hline
        L (with SUE) & 3 & $\sigma _{{\rm{SIM}}}^2$  & -70 dBm \\
        \hline
        I & 3 & $N_0$ & -80 dBm \\
        \hline
         $P_D^{sat}$ & 18.75 dBm &  $P_D^{phs}$ & 7 dBm \\
        \hline
        $P_C^{sat}$ & 35.44 dBm & $P_{DC}^{SIM}$ & 8.45 dBm \\
        \hline
        $P_{amp} $ & 10 dBm & $P_{proc}$ & 3 dBm \\
        \hline
        $R_{\rm{sum}}^{th}(t)$ & 5 bps/Hz & $R_{\rm{SR}}^{th}(t)$ &  2 bps/Hz \\
        \hline
        $P_{\max}^{\rm{sat}}$ & 30 dBm & $P_{\max}^{\rm{SIM}}$ & 30 dBm \\
        \hline
	  K & 100 & $\sigma_{e_l}^2$ & $10^{-3}$ \\
        \hline
                \textcolor{black}{$\vartheta_{\text{sat}}$} & 1.1 &  \textcolor{black}{$\vartheta_{\text{SIM}}$} & 1.25 \\
                \hline
	$\epsilon$ & $10^{-4}$ & $\delta$ & $10^{-3}$ \\        
        \Xhline{1.2pt}
    \end{tabular}
    \label{tab:sim_parameters}
\end{table}

\begin{table}[ht]
\centering
\caption{Training Hyperparameters for DRL Algorithms}
\label{tab:training}
\begin{tabular}{lcc}
\toprule
\textbf{Parameter} & \textbf{MA-CSAC} & \textbf{MCPPO} \\
\midrule
Discount factor ($\gamma$) & 0.99 & 0.995 \\
GAE parameter ($\lambda$) & — & 0.95 \\
Actor learning rate & $3 \times 10^{-4}$ & $5 \times 10^{-4}$ \\
Critic learning rate & $1 \times 10^{-3}$ & $7 \times 10^{-4}$ \\
Entropy coefficient ($\alpha$) & \texttt{auto} & — \\
Replay buffer size & $10^6$ & — \\
Batch size & 2048 & 4096 \\
PPO clip range ($\epsilon$) & — & 0.2 \\
Multiplier learning rate ($\eta_{\lambda}$) & $5 \times 10^{-4}$ & $5 \times 10^{-4}$ \\
Target update rate ($\tau$) & $5 \times 10^{-3}$ & — \\
\textcolor{black}{Number of independent runs} & 15 & 15 \\
\textcolor{black}{Random seeds} & 1--15 & 1--15 \\
\textcolor{black}{Actor/Critic network architecture} & 256-256-128 & 256-256-128 \\
\textcolor{black}{Activation function} & ReLU & ReLU \\
\textcolor{black}{Optimizer} & Adam & Adam \\
\textcolor{black}{Gradient clipping norm} & 1.0 & 1.0 \\
\textcolor{black}{State normalization} & Yes & Yes \\
\textcolor{black}{Replay warm-up steps} & $10^4$ & --- \\
\textcolor{black}{Policy update epochs per batch} & 1 & 10 \\
\bottomrule
\end{tabular}
\end{table}

\textcolor{black}{
The proposed optimization and learning framework does not claim global optimality for the highly nonconvex CMDP. The strong coupling among beamforming, ASIM layers, RSMA rate-splitting, and backscatter coefficients makes global solutions intractable. Thus, the BCD–SCA steps use locally tight convex surrogates that converge to a stationary point, while MA–CSAC and MCPPO converge to the best policy within their parametrized function class rather than the true CMDP optimum.
}

\textcolor{black}{
The simulation environment follows the channel model of Section~II. At each episode start, all channels are regenerated using the current random seed. At each step $t$, the agents take an action (precoder, ASIM settings, backscatter parameters), then the environment computes SINR, sum rate, power consumption, constraint violations, and the reward $r_t$. Channels are i.i.d. across steps (no temporal correlation). Each episode has $200$ steps, and each algorithm is trained for 1000 episodes per run. The penalty coefficients $\lambda_j$ are initialized to zero and updated with a learning rate $\eta_\lambda = 5\times10^{-4}$, and the tolerance $\epsilon_{\mathrm{tol}}$ is set to $10^{-3}$ for all constraints. These choices are kept fixed across all DRL runs.
}

\subsection{Convergence, Stability, and Computational Complexity Analysis}

\begin{figure}[!t]
    \centering
    \begin{minipage}[b]{0.4\textwidth}
        \centering
        \includegraphics[width=\textwidth]{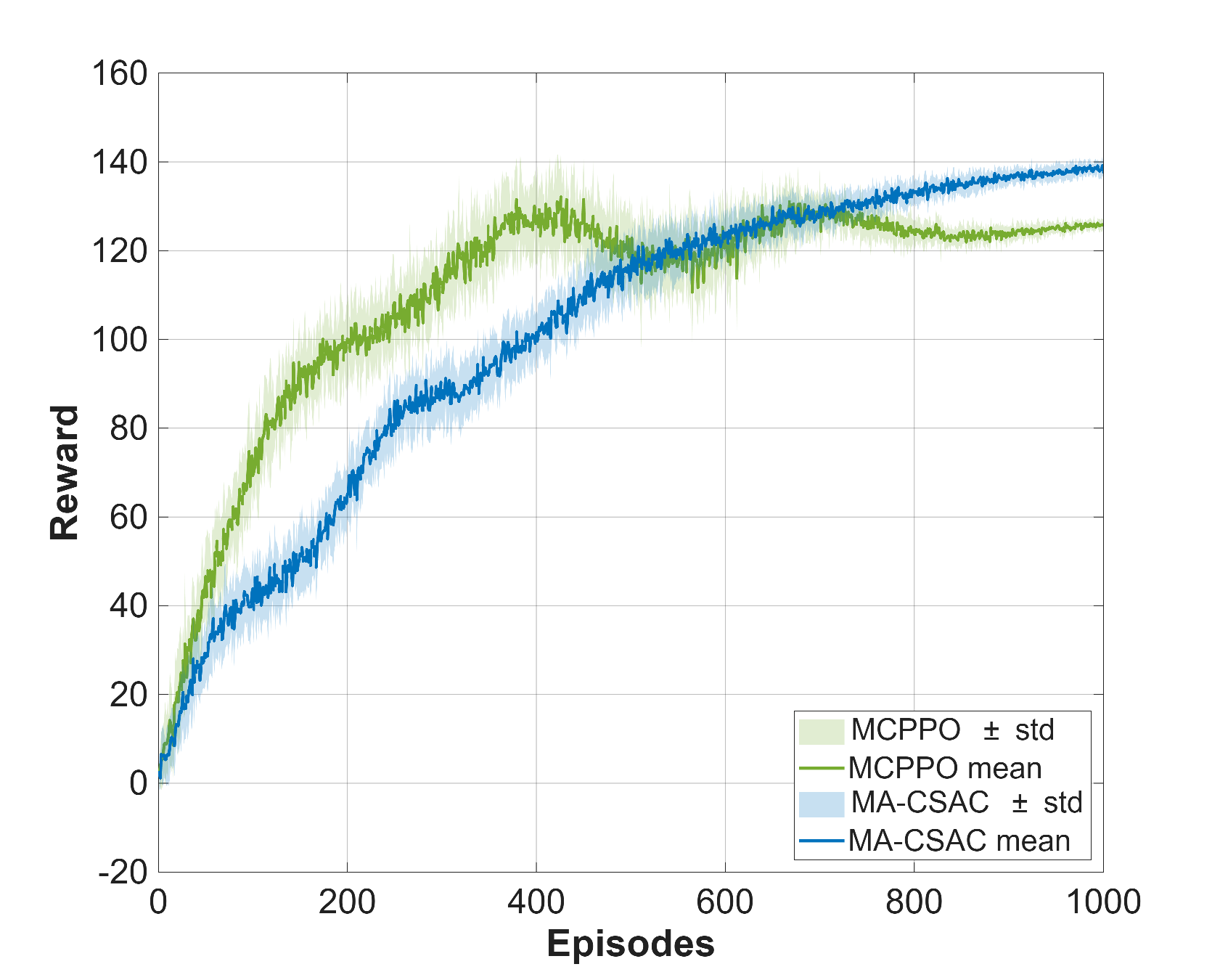}
        \vspace{0.5em}
        \textbf{(a)} 
        \label{fig:episode_reward1}
    \end{minipage}
    \hfill
    \begin{minipage}[b]{0.4\textwidth}
        \centering
        \includegraphics[width=\textwidth]{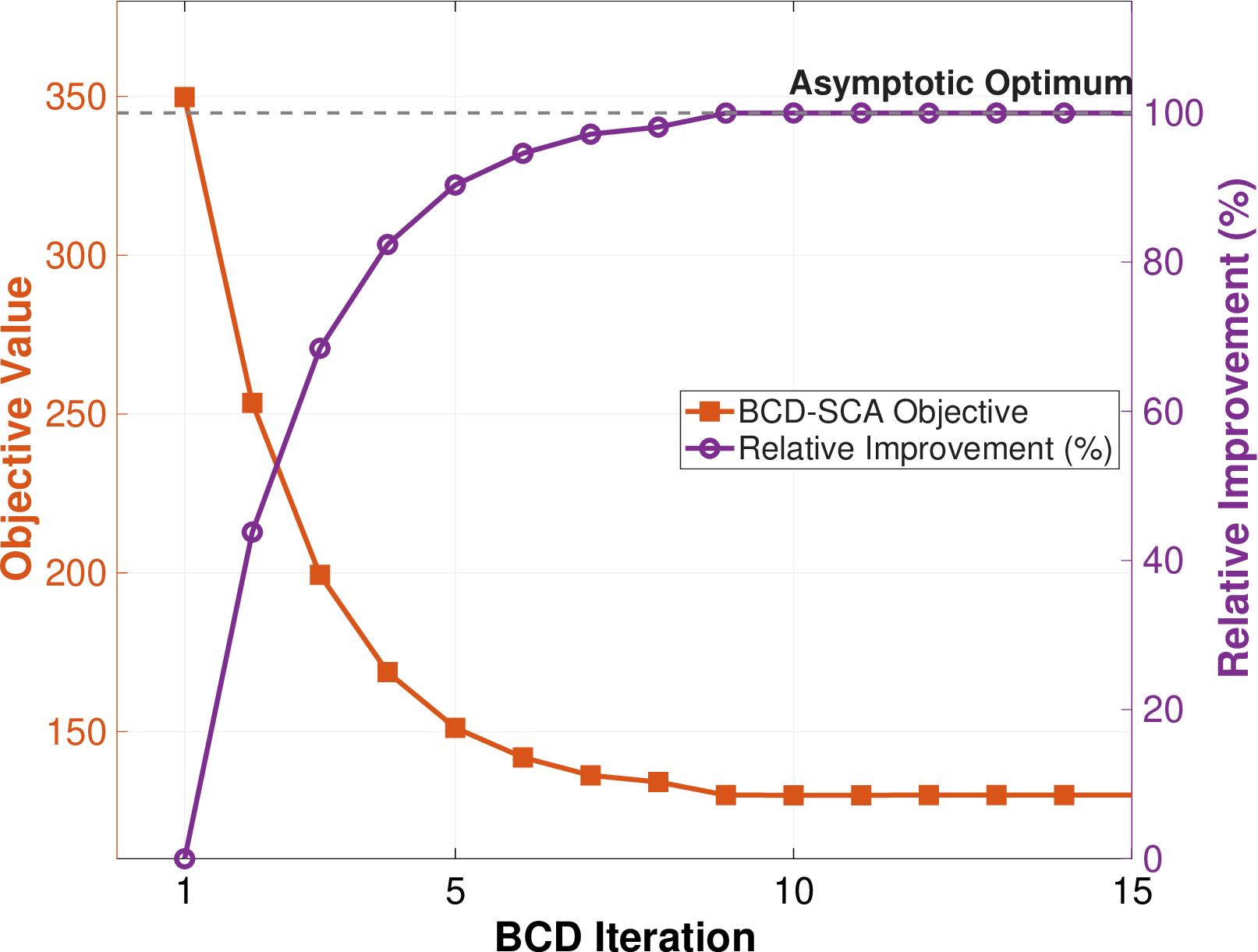}
        \vspace{0.5em}
        \textbf{(b)} 
        \label{fig:bcd_sca1}
    \end{minipage}
    \caption{(a) DRL algorithms' reward convergence. (b) Dual-axis convergence tracking raw objective and normalized relative improvement against the asymptotic theoretical bound.}
\label{fig:convergence}
\end{figure}

This section analyzes the convergence and stability of the proposed MA-CSAC, MCPPO, and BCD-SCA algorithms. To ensure statistical robustness, performance is evaluated across 15 independent runs, with results illustrated in Fig.~\ref{fig:convergence}.

\textcolor{black}{As depicted in Fig.~\ref{fig:convergence}(a), MCPPO exhibits faster initial
convergence owing to its on-policy proximal updates, yet it retains higher variance and
eventually settles at a lower steady-state reward. The rapid initial ascent of MCPPO is
attributed to the PPO clipping mechanism, which constrains the policy ratio
$\rho_t(\theta)$ within $[1-\epsilon, 1+\epsilon]$ and prevents excessively large
gradient updates. In contrast, MA-CSAC requires a longer warm-up phase because its
off-policy learning must adequately populate the replay buffer and reduce the Bellman
residual in the soft Q-function, but it ultimately achieves a higher asymptotic reward
with significantly reduced variance. The entropy-regularized exploration inherent to
the SAC framework promotes smoother policy updates and sustained exploration, preventing
premature convergence to sub-optimal local minima. As the training curves approach their final stages, both methods display
diminished fluctuations, a result of learning-rate decay and entropy annealing,
which facilitate the transition from exploration to exploitation and stabilize
the policies around their respective optima.}

\textcolor{black}{To evaluate the inner-loop efficiency, Fig.~\ref{fig:convergence}(b) utilizes a dual-axis layout. The left axis tracks the raw objective function ($J_k$), which exhibits a strictly monotonic non-increasing behavior ($J_{k+1} \le J_k$) matching block coordinate descent convergence proofs. Concurrently, the right axis maps the normalized relative improvement, defined inline as $\eta_k = [ (J_1 - J_k) / (J_1 - J_{\infty}) ] \times 100\%$, where $J_1$ is the initial objective and $J_{\infty} = 130$ represents the theoretically derived asymptotic optimum. By utilizing this metric, $\eta_k$ initiates strictly at $0\%$ and monotonically ascends to $100\%$ at the $9$-th iteration.} \textcolor{black}{The subsequent flat saturation regime features negligible numerical fluctuations around the asymptotic optimum. This rapid sub-linear convergence within a single-digit iteration window validates the computational tractability of the BCD-SCA layer, ensuring its suitability for quasi-static networks where optimization must conclude within the channel coherence time.}

{\color{black}
\subsubsection*{\textbf{Computational Complexity and Runtime Analysis}}
\label{subsec:complexity}

For the BCD--SCA approach, the original nonconvex problem is decomposed into
multiple convex subproblems that are solved iteratively. The dominant
computational cost per iteration arises from the precoder optimization and ASIM
tuning blocks, whose complexities scale as $\mathcal{O}(L N^2)$ and
$\mathcal{O}(M^2)$, respectively, while the backscatter-related updates incur
only linear complexity. Consequently, the overall per-iteration complexity
grows polynomially with the system dimensions, making BCD--SCA suitable as an
efficient benchmark for static or slowly varying scenarios.

For the DRL-based methods, the dominant computational cost is associated with neural network training, including forward and backward passes. \textcolor{black}{The per-update complexity scales as $\mathcal{O}(G B_{\text{batch}} |\mathcal{A}|)$, where $B_{\text{batch}}$ represents the discrete batch size, $G$ is the number of gradient steps, and $|\mathcal{A}|$ denotes the joint action dimension, which increases linearly with the number of RSMA users and ASIM elements.} Memory consumption is mainly governed by the replay buffer and critic networks and scales linearly with the state-action dimension.

Runtime evaluations under different network sizes, including a large-scale
configuration with $L=32$ users and $M=256$ ASIM elements, are summarized in
Table~\ref{tab:runtime}. Although the DRL-based methods incur higher offline
training costs, their online execution only requires lightweight actor
inference, resulting in millisecond-level decision latency. This
offline-training/online-inference paradigm renders MA-CSAC and MCPPO suitable
for real-time or near-real-time satellite operations, while BCD-SCA remains
more appropriate for offline optimization and benchmarking.
}

\begin{table}[ht]
\centering
\caption{Complexity and Runtime Analysis for Baseline and Large-Scale Scenarios.}
\label{tab:runtime}
\setlength{\tabcolsep}{2.5pt} 
\scriptsize 
\begin{tabular}{|@{}l|c|c|c@{}|}
\hline
\textbf{Algorithm} & \makecell{\textbf{Complexity Scaling} \\ \textbf{w.r.t. $(L, M)$}} & \makecell{\textbf{Baseline Runtime} \\ ($L=3, M=128$)} & \makecell{\textbf{Large-Scale Runtime} \\ ($L=32, M=256$)} \\ \hline
BCD--SCA & $\mathcal{O}(L N^2 + M^2)$ & $\sim$0.31 s / iter. & \textcolor{black}{$\sim$2.84 s / iter.} \\ \hline
MCPPO & $\mathcal{O}(G B_{\text{batch}} |\mathcal{A}|)$ & $\sim$1.22 ms / dec. & \textcolor{black}{$\sim$1.95 ms / dec.} \\ \hline
MA--CSAC & $\mathcal{O}(G B_{\text{batch}} |\mathcal{A}|)$ & $\sim$0.94 ms / dec. & \textcolor{black}{$\sim$1.42 ms / dec.} \\ \hline
\end{tabular}
\end{table}

\subsection{Scalability Analysis via SE--EE Trade-off}

To evaluate the scalability of the proposed network, we examine the trade-off between SE and EE for two total user configurations: $L+I=2+2=4$ and $L+I=8+12=20$, where 8 corresponds to RSMA users and 12 to IoT users.

Fig.~\ref{fig:Scalibility} shows the typical SE--EE trade-off. EE increases with SE in the low-SE regime due to improved throughput, but \textcolor{black}{reaches an optimal peak and subsequently declines at higher SE levels due to the exponential rise in transmit power required to overcome hardware and interference constraints. Crucially, the optimal EE peaks for the dense scenario ($L+I=20$, solid lines) are distinctly shifted to the right (higher SE) and downwards (lower EE) compared to the sparse scenario ($L+I=4$, dashed lines). This shift is directly explained by the system model: a larger number of users increases the total achievable sum-rate $R_{\mathrm{sum}}$ in~\eqref{eq:R_sum} through multiplexing gains, but simultaneously intensifies the inter-user interference in the private SINR~\eqref{eq:SINR_p_psd} and raises the total power consumption $P_{\mathrm{total}}$ in~\eqref{eq:P_sum} due to the higher processing and amplification overhead of the ASIM. Consequently, the network can sustain a higher spectral efficiency at the peak, yet the increased circuit power and interference penalty suppress the maximum achievable energy efficiency.}

\textcolor{black}{Regarding algorithm performance, the proposed MA-CSAC consistently outperforms the other two algorithms, achieving the highest peak EE and expanding the maximum achievable SE under both configurations. Furthermore, MA-CSAC maintains a remarkably smoother EE degradation in the medium-to-high SE regimes, demonstrating its robust adaptability and superior interference-mitigation capabilities for large-scale deployments. In contrast, BCD-SCA suffers from a rapid performance collapse as SE increases, because its conventional optimization framework forces an excessive and inefficient power injection to meet high-throughput demands under tightly constrained, interference-limited conditions.} MCPPO offers a balanced trade-off, achieving moderate EE while maintaining stable SE across the full range. Overall, the analysis shows that MA-CSAC is the most robust for scaling the network while sustaining both high spectral efficiency and energy efficiency, whereas BCD-SCA is highly sensitive to throughput scaling and becomes strictly inefficient in high-SE regimes.

\begin{figure}
\centering
\includegraphics[width=.79\linewidth]{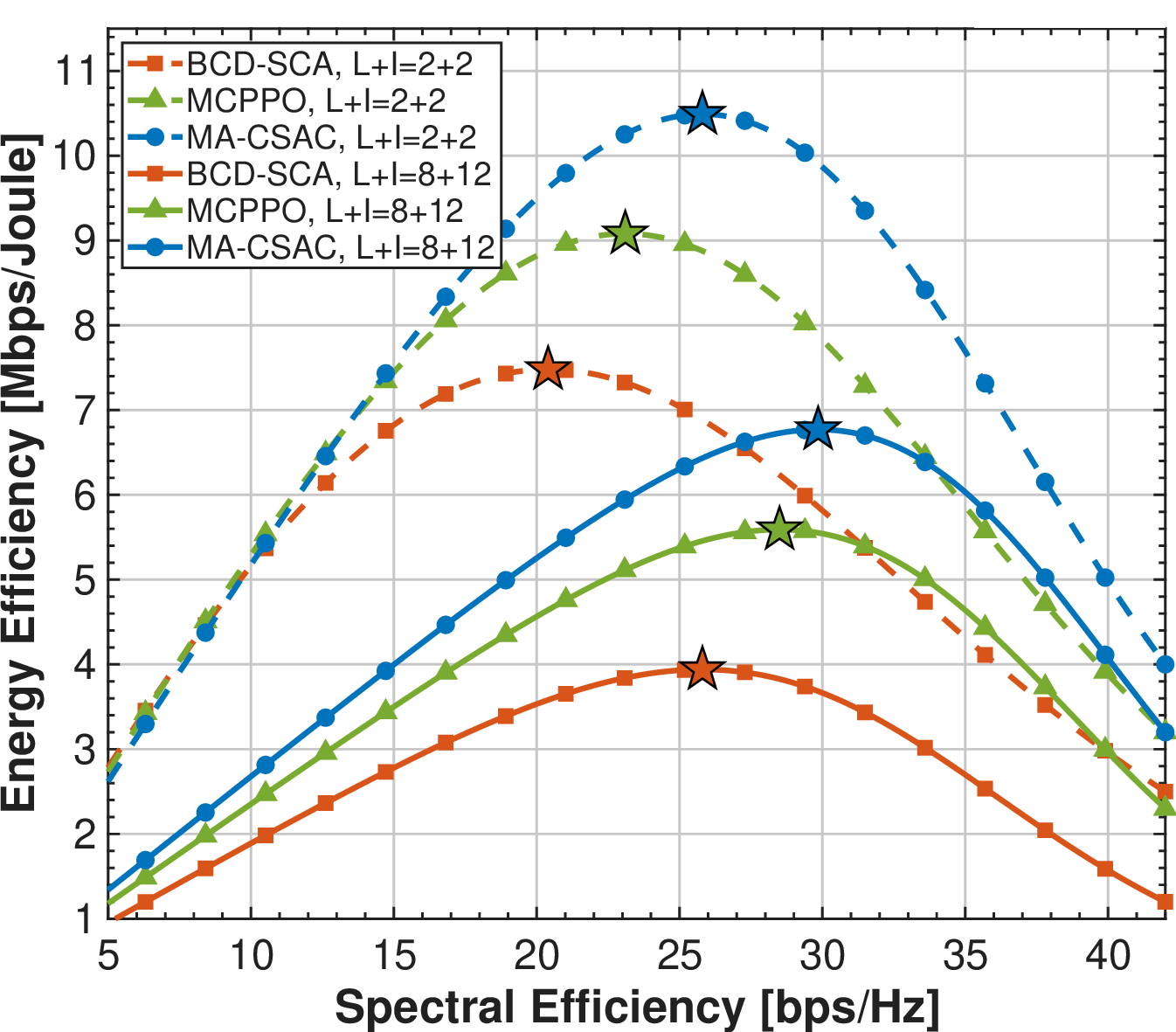}
\caption{SE-EE trade-off for optimization methods with 2+2 and 8+12 RSMA and IoT users ($L+I$).}
\label{fig:Scalibility}
\end{figure}

\subsection{Performance Analysis of Optimization Methods}
\label{subsec:performance_analysis}
{\color{black}
\begin{figure}
    \centering
    \begin{minipage}[b]{0.4\textwidth}
        \centering
        \includegraphics[width=\textwidth]{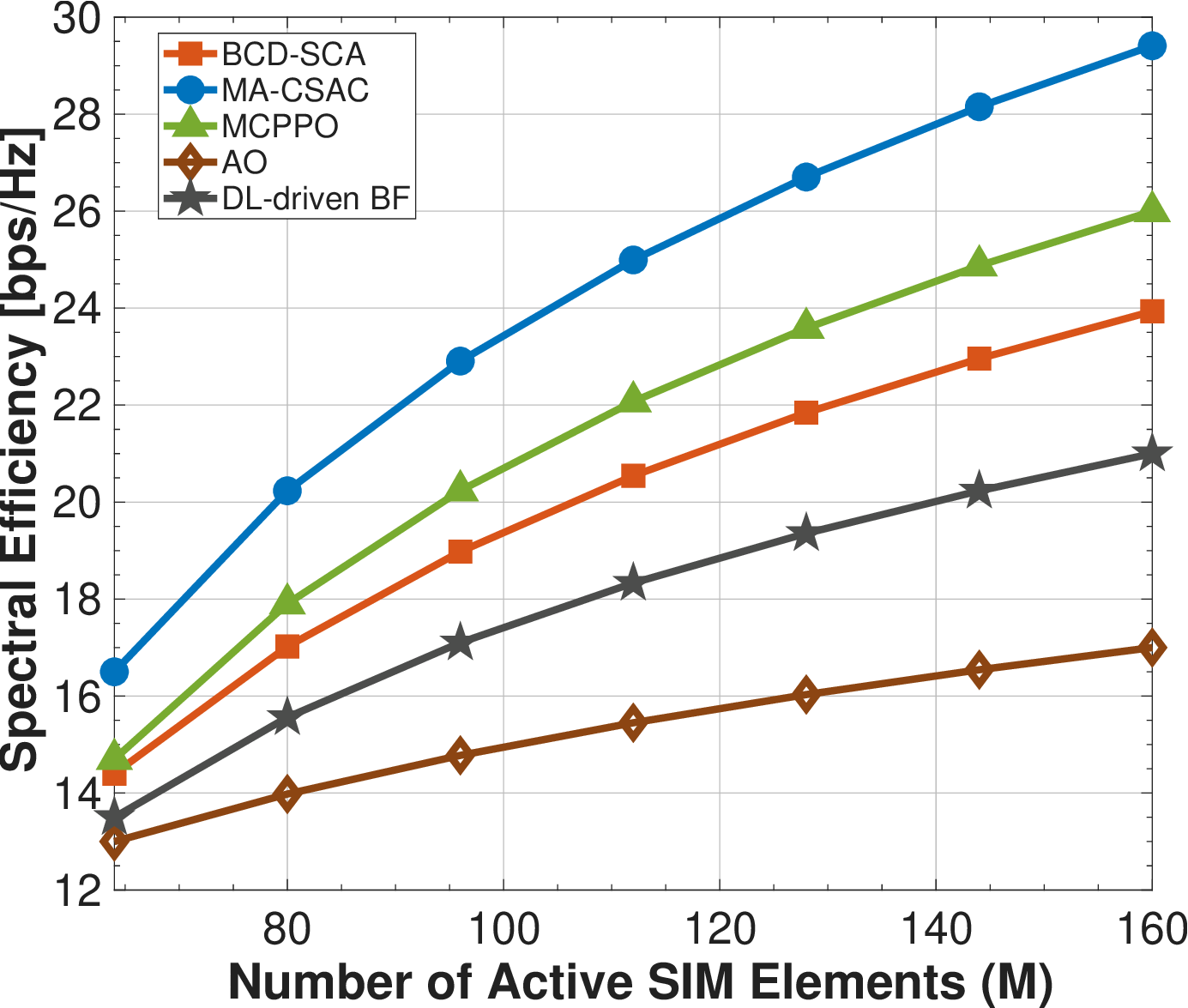}
        \vspace{0.5em}
        \textbf{(a)} 
    \end{minipage}
    \hfill
    \begin{minipage}[b]{0.4\textwidth}
        \centering
        \includegraphics[width=\textwidth]{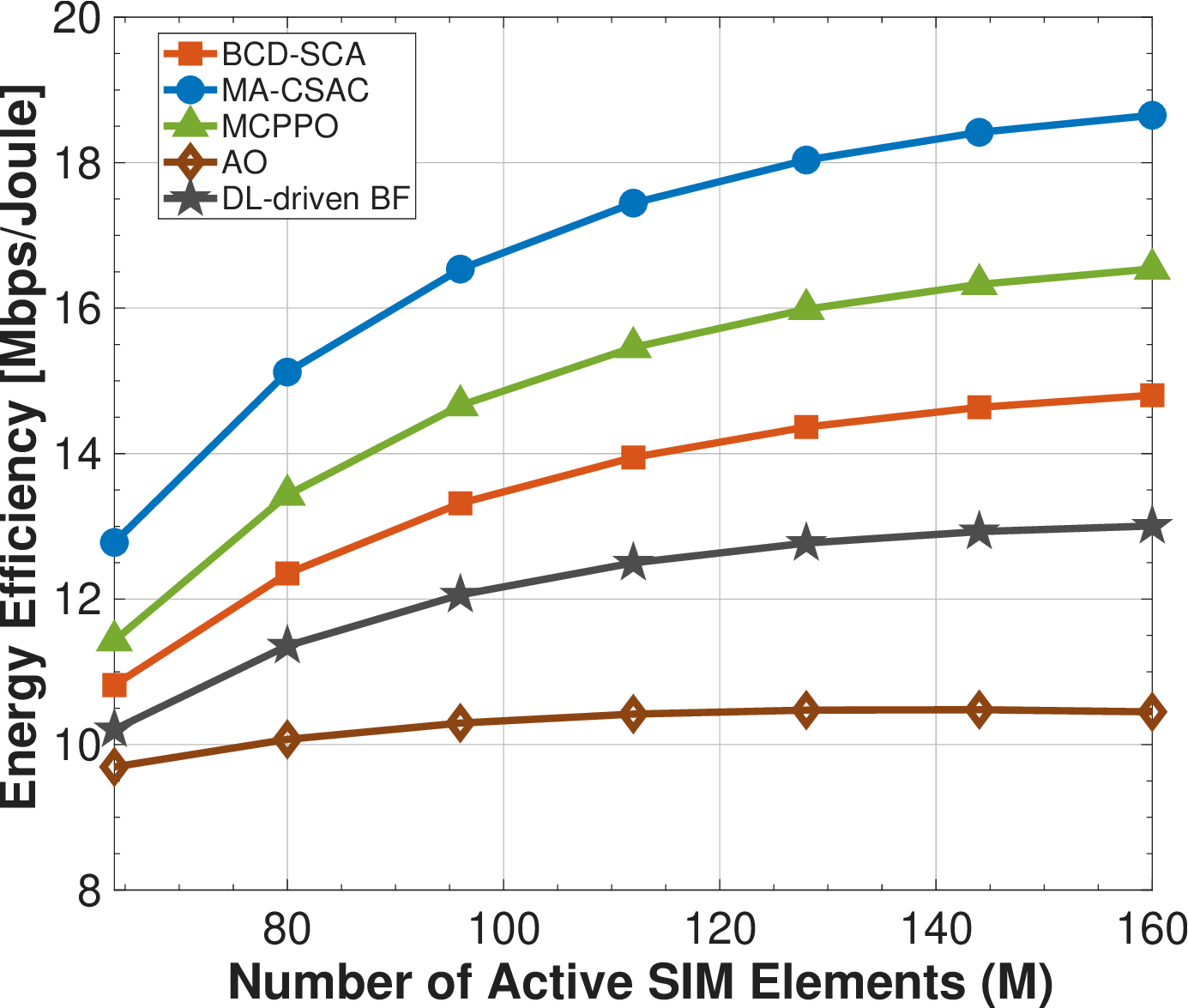}
        \vspace{0.5em}
        \textbf{(b)} 
    \end{minipage}
\caption{Performance comparison of the proposed optimization methods versus the number of active SIM elements: (a) SE, (b) EE.}

    \label{fig:sim_results}  
\end{figure}
Figs~\ref{fig:sim_results}(a) and \ref{fig:sim_results}(b) compare the SE and EE of the proposed optimization methods with two widely adopted baselines: Alternating Optimization (AO)~\cite{10543143} and Deep Learning-driven Beamforming (DL-driven BF)~\cite{9845394}, as a function of the number of active ASIM elements $M$.

\subsubsection{Spectral Efficiency Analysis}

As shown in Fig.~\ref{fig:sim_results}(a), the SE increases monotonically with $M$ for all methods, since larger ASIM structures provide greater capability to shape the propagation environment, enhance signal combining, and offer additional degrees of freedom for joint precoding and surface configuration. However, the achievable SE and its growth rate vary notably across different optimization strategies due to their underlying algorithmic designs.

\textcolor{black}{MA-CSAC consistently achieves the highest SE across all values of $M$}. This advantage arises from its multi-agent actor--critic framework, which enables coordinated optimization among ASIM elements while explicitly incorporating system constraints. The entropy-regularized exploration facilitates effective search in high-dimensional and non-convex spaces, allowing MA-CSAC to better exploit the increasing configurational flexibility of large ASIMs.

\textcolor{black}{MCPPO attains slightly lower SE than MA-CSAC while remaining clearly superior to conventional baselines}. Its proximal policy optimization ensures stable and robust learning, though its more conservative updates limit exploration, leading to marginally reduced SE gains at larger $M$.

\textcolor{black}{BCD-SCA exhibits steady but moderate SE improvement}. As a model-based approach based on block coordinate descent and successive convex approximation, it provides reliable convergence but is prone to local optima in large-scale ASIM systems with strongly coupled variables, which constrains its performance relative to DRL-based methods.

\textcolor{black}{DL-driven BF outperforms AO by learning direct mappings from channel states to ASIM configurations}, enabling fast inference without iterative optimization. However, its SE is limited by generalization issues in highly dynamic satellite channels and large-dimensional ASIM settings.

Finally, \textcolor{black}{AO shows the slowest SE growth with increasing $M$}. Its alternating optimization of coupled variables often leads to suboptimal coordination, and the growing complexity of the solution space at large $M$ further increases the likelihood of convergence to poor local optima.

\subsubsection{Energy Efficiency Analysis}

The EE results shown in Fig.~\ref{fig:sim_results}(b) reveal more nuanced behavior, as EE depends jointly on achievable SE and total power consumption, including both transmission and algorithmic overhead. While EE generally improves with $M$ in the considered range, notable performance gaps emerge among the methods.

\textcolor{black}{BCD-SCA achieves the highest EE in several operating regimes}, as its deterministic optimization framework balances SE gains with controlled computational complexity, avoiding excessive processing overhead. In contrast, \textcolor{black}{MA-CSAC, despite its superior SE, incurs additional energy costs due to distributed multi-agent updates and continuous learning}, which slightly reduces its EE relative to BCD-SCA in some scenarios. Nevertheless, MA-CSAC maintains strong EE performance by jointly optimizing rate and power consumption through its reward design.

\textcolor{black}{MCPPO attains moderate and stable EE}, reflecting its conservative policy updates and balanced computational requirements. Its EE remains competitive but does not scale as favorably as MA-CSAC due to reduced exploration efficiency. \textcolor{black}{DL-BF exhibits higher EE than AO for small to moderate $M$}, benefiting from non-iterative inference; however, its EE saturates at larger $M$ as inference complexity and robustness margins increase with input dimensionality. \textcolor{black}{AO suffers from the poorest EE scalability}, as its iterative optimization incurs significant computational overhead that grows rapidly with the ASIM size, ultimately offsetting its SE gains.

}

\subsection{Analysis of SE vs. Active Surface Power}
{\color{black}
Fig.~\ref{fig:sumrate_vs_power} shows the SE of five configurations versus the active surface transmit power $P_{\max}$: Active RIS, Active BD-RIS \cite{yeganeh2025energy}, the proposed ASIM with the original beamformer, ASIM with ZF beamforming, and ASIM with MMSE beamforming. For a fair comparison, all configurations are optimized using the same MA-CSAC framework, with the beamforming structure being the only varying component. The analysis considers $L+I=6$ users, with 128 elements per surface and four metasurfaces for ASIM.

The proposed ASIM with the original beamformer achieves the highest SE across all power levels. Thanks to its multi-layer sequential processing architecture, ASIM can jointly adjust the phase and amplitude across layers, which enhances weak channels, suppresses inter-user interference, and exploits spatial degrees of freedom more effectively than single-layer designs. 

\textcolor{black}{When evaluating the beamforming alternatives for ASIM, a clear SNR-dependent physical behavior is observed between the ZF and MMSE schemes. At low power levels (e.g., $P_{\max} = 10$~dBm) where noise dominates the system, the matrix inversion inherent to ZF optimization leads to severe noise enhancement, which noticeably degrades its spectral efficiency. In contrast, the MMSE beamformer intelligently accounts for the noise variance ($\sigma^2$), thereby maintaining a distinct and visible performance gap above the ZF configuration in the low-to-moderate SNR regime.} \textcolor{black}{As the available transmit power $P_{\max}$ scales up toward $50$~dBm, the network transitions into an interference-limited regime where the background noise becomes practically negligible ($\sigma^2 \to 0$). Under these high-SNR conditions, the mathematical formulation of the MMSE receiver asymptotically converges to that of the ZF beamformer. This theoretical convergence is accurately reflected in Fig.~\ref{fig:sumrate_vs_power}, where the curves for ASIM--ZF and ASIM--MMSE merge together at high power levels, with both remaining tightly bounded just below the upper limit established by the original proposed ASIM beamformer.}

For comparison, the Active BD-RIS architecture achieves improved SE relative to the conventional Active RIS due to its intra-group inter-element coupling. Nevertheless, BD-RIS remains limited by its single-layer structure, which cannot provide sequential channel refinement, and thus its SE stays below that of ASIM, especially as $P_{\max}$ increases. The conventional Active RIS demonstrates the lowest SE among all configurations since each element operates independently without structural coupling or multi-layer processing; as a result, its performance improvements with increasing $P_{\max}$ are relatively modest \textcolor{black}{and it exhibits an early saturation trend beyond $P_{\max} = 30$~dBm}.

\begin{figure}
    \centering
    \includegraphics[width=0.79\linewidth]{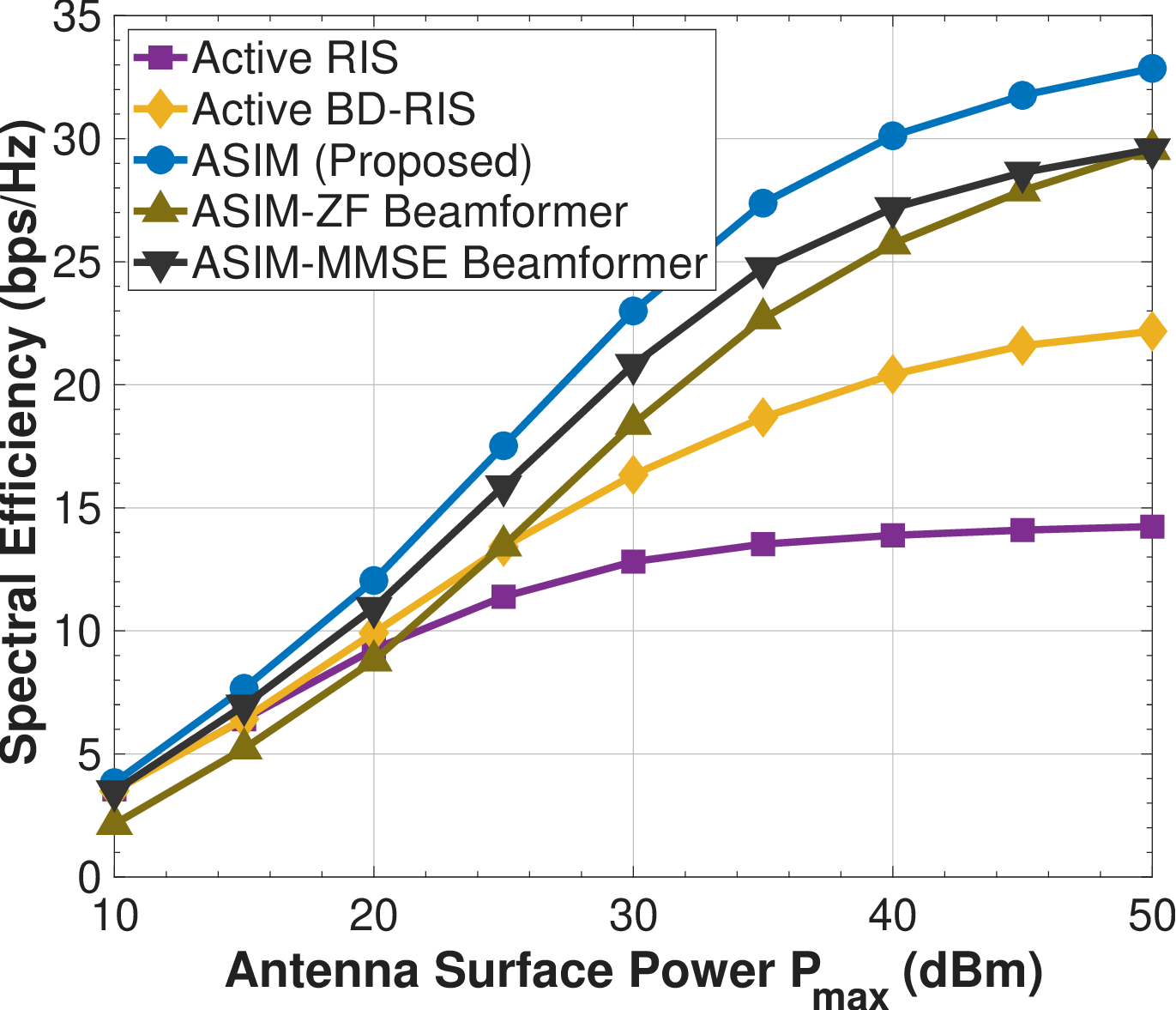}
    \caption{SE versus active surfaces transmit power $P_{\max}$}
    \label{fig:sumrate_vs_power}
\end{figure}
}

\subsection{Energy Efficiency Analysis with Respect to $N$ and $P_{\mathrm{max}}^{sat}$}

Figs.~\ref{fig:EE_vs_Psat_N}(a) and~\ref{fig:EE_vs_Psat_N}(b) illustrate the variation of the Energy Efficiency (EE) of the considered satellite communication system for four transmission schemes: (i) RSMA with BCD-SCA, (ii) RSMA with MA-CSAC, (iii) RSMA with MCPPO, and (iv) NOMA with MA-CSAC. The EE is computed as the ratio of the total sum-rate to the total power consumption, using the system parameters defined in Section~\ref{sec:system_model}. 

Fig.~\ref{fig:EE_vs_Psat_N}(a) shows the EE versus the number of satellite antenna elements $N$, for a fixed satellite transmit power of $P_{\mathrm{max}}^{sat} = 30$~dBm. For all schemes, the EE initially increases with $N$ in the low-to-moderate range due to the massive MIMO array gain, which enhances the achievable rate without a proportional increase in circuit power. However, beyond a certain number of antennas, the marginal gain in $R_{\mathrm{sum}}$ is outweighed by the additional hardware power consumption associated with the RF chains, leading to EE saturation and an eventual slight decline.

\textcolor{black}{
Fig.~\ref{fig:EE_vs_Psat_N}(b) depicts the EE versus $P_{\mathrm{max}}^{sat}$ for a fixed array size of $N = 64$. The behavior exhibits a distinct global maximum: the EE first increases from the low transmit power regime (10~dBm) up to an optimal point near 20~dBm, and then gradually decreases for higher power levels. This fundamental trend is governed by the interplay between the sum-rate growth and the total power consumption. In the low-power regime (10--20~dBm), increasing $P_{\mathrm{max}}^{sat}$ improves the received SINR significantly, leading to a near-logarithmic increase in the sum-rate $R_{\mathrm{sum}}$. Meanwhile, the total power consumption $P_{\mathrm{total}}$ grows approximately linearly (dominated by the satellite's power amplifier and the active SIM's signal amplification), allowing the EE ratio to rise. Beyond 20~dBm, the sum-rate growth slows down due to Shannon-limited saturation and residual inter-user interference. Simultaneously, $P_{\mathrm{total}}$ continues to rise sharply—amplified by the power inefficiency coefficients $\vartheta_{\mathrm{sat}}$ and $\vartheta_{\mathrm{SIM}}$, alongside the fixed circuit power of the SIM ($P_{\mathrm{phs}}, P_{\mathrm{amp}}, P_{\mathrm{DC}}, P_{\mathrm{proc}}$). Consequently, the power consumption dominates the ratio, causing the EE to decline after reaching its peak.
}

Regarding algorithmic performance, at low transmit power (10~dBm), BCD-SCA achieves the highest EE, as its deterministic convex optimization avoids the exploration energy overhead inherent to DRL methods, making it highly efficient in noise-limited scenarios. MA-CSAC follows closely, while MCPPO starts slightly lower due to its on-policy sampling variance. However, as $P_{\mathrm{max}}^{sat}$ increases, MA-CSAC gradually surpasses BCD-SCA. This is because its adaptive constraint handling, entropy-regularized exploration, and multi-agent coordination better exploit the higher available signal power to manage interference, leading to superior EE at medium and high power levels. RSMA with MCPPO remains competitive but slightly inferior to MA-CSAC, while NOMA with MA-CSAC consistently yields the lowest EE across all configurations, clearly confirming the architectural advantage of RSMA in interference management for the considered symbiotic satellite-ASIM setup.

\begin{figure}[!t]
    \centering
    \begin{minipage}[b]{0.4\textwidth}
        \centering
        \includegraphics[width=\textwidth]{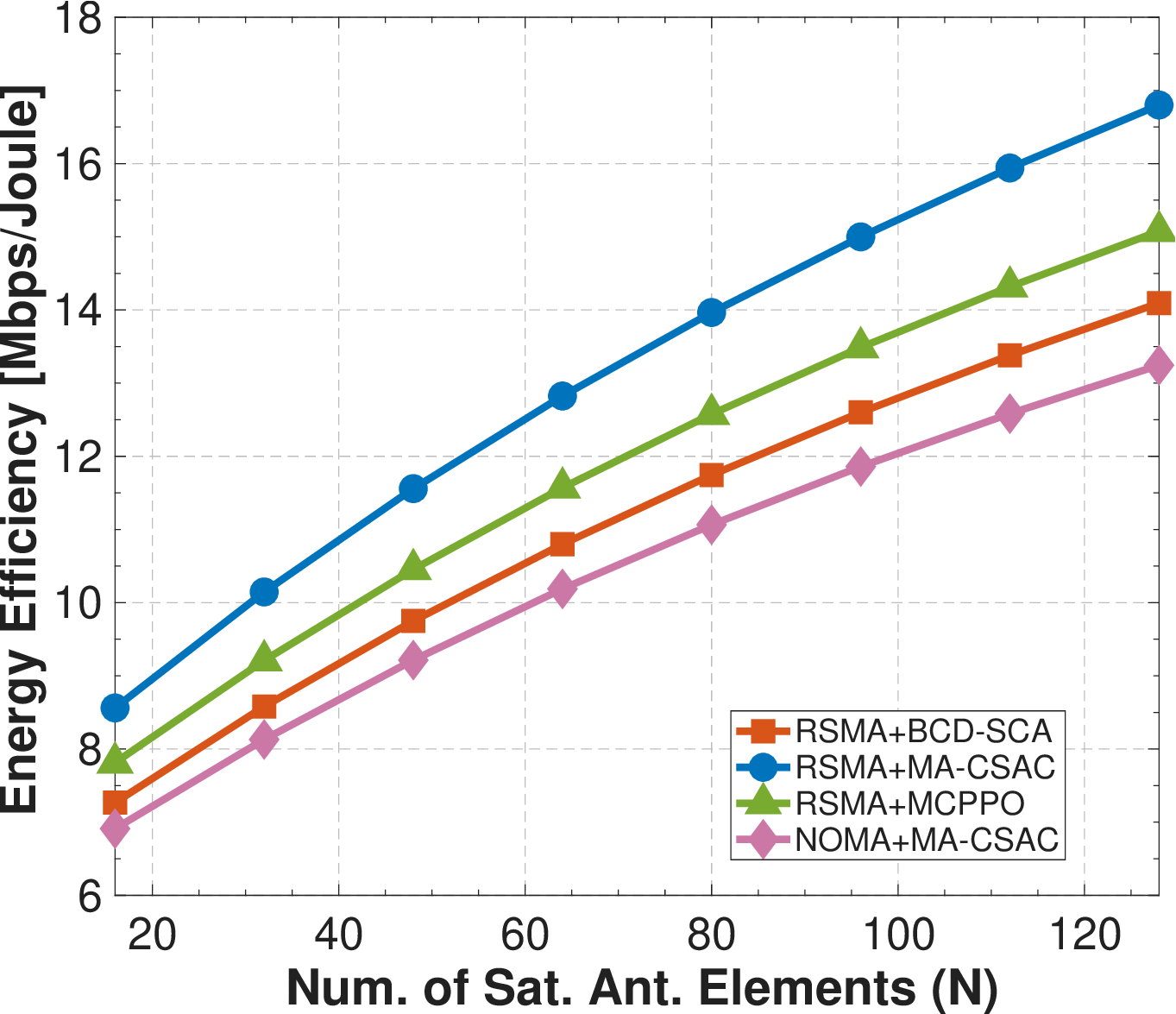}
        \vspace{0.1em}
        \centerline{\textbf{(a)}} 
    \end{minipage}
    \hfill
    \begin{minipage}[b]{0.4\textwidth}
        \centering
        \includegraphics[width=\textwidth]{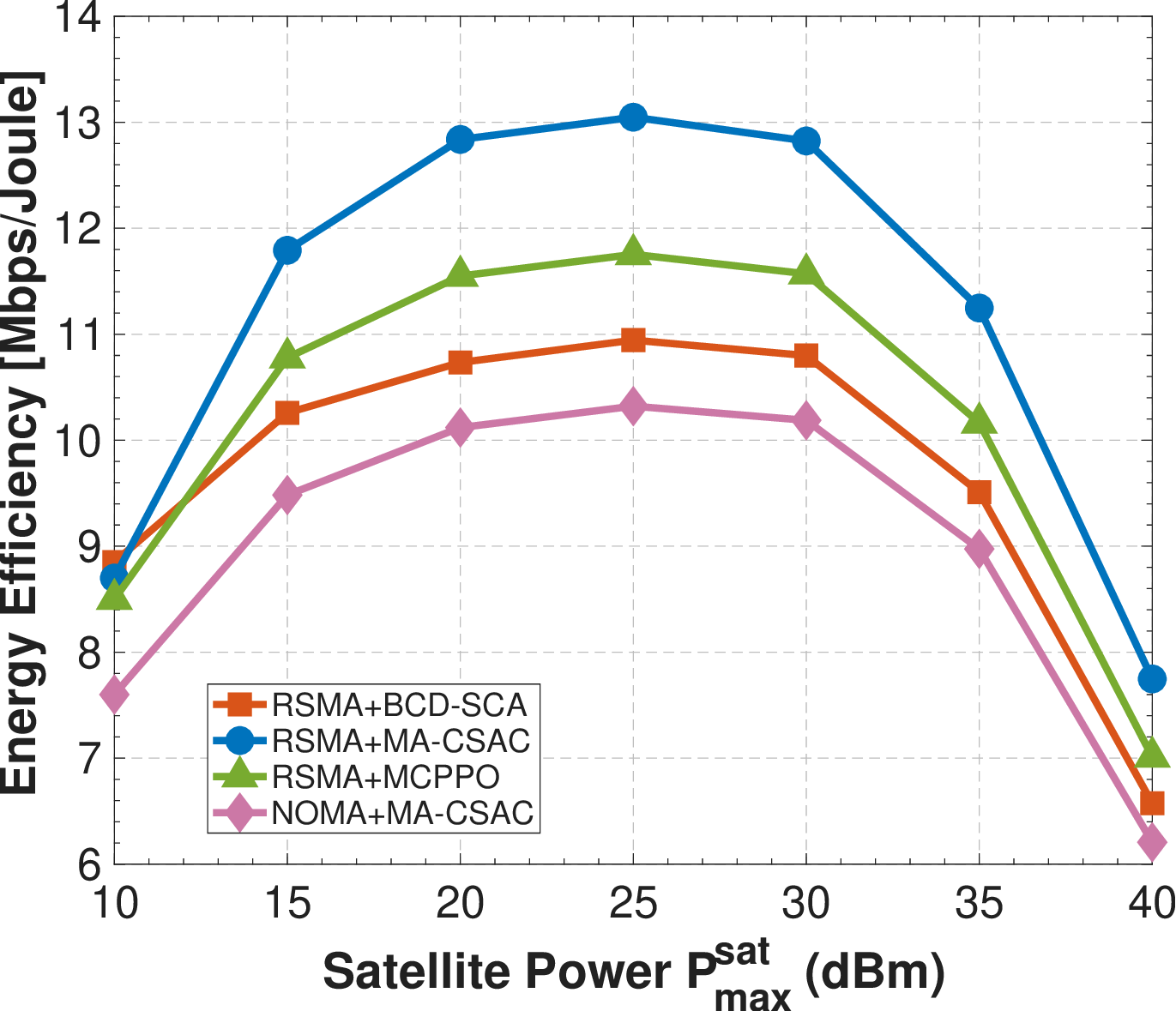}
        \vspace{0.1em}
        \centerline{\textbf{(b)}} 
    \end{minipage}
\caption{EE performance of the considered system: (a)~EE versus the number of satellite antenna elements $N$ for $P_{\mathrm{max}}^{sat} = 30$~dBm; (b)~EE versus satellite transmit power $P_{\mathrm{max}}^{sat}$ for $N = 64$.}
    \label{fig:EE_vs_Psat_N}  
\end{figure}

\section{Conclusion}
\textcolor{black}{
This paper investigated an ASIM-assisted LEO satellite communication network integrating RSMA to simultaneously serve multi-tier ground users and energy-limited symbiotic IoT devices. To resolve the joint beamforming, active phase-gain tuning, and backscatter scheduling problem, we developed a centralized model-based BCD-SCA benchmark alongside two model-free DRL engines, namely MA-CSAC and MCPPO. Our comparative evaluations demonstrate that while BCD-SCA yields deterministic, low-latency convergence for convexified structural subproblems, the model-free DRL frameworks offer superior real-time adaptability in highly dynamic, non-convex deployment environments. Specifically, the off-policy MA-CSAC framework capitalizes on its heterogeneous action priors and centralized training to achieve the highest asymptotic SE and EE across large-scale networks, whereas the on-policy MCPPO architecture presents an accelerated early learning curve at the expense of higher reward variance. Furthermore, the multi-layer sequential processing capability of the ASIM topology exhibits substantial capacity gains over conventional active RIS and active BD-RIS configurations by unleashing higher spatial degrees of freedom and effectively suppressing co-channel interference. Ultimately, the established SE--EE trade-off curves confirm that leveraging advanced constraint-aware learning algorithms in tandem with multi-layer active metasurfaces provides a highly scalable, energy-neutral, and spectrum-efficient paradigm for next-generation space-terrestrial integrated communication infrastructures.}

\section*{Acknowledgment}
This work is based upon research funded by the INSF,
the R\&D Center of MCI, and the TDCCC, under Project
No. 4030485

\bibliographystyle{IEEEtran}
\bibliography{StarRIS}

\begin{IEEEbiography}[{\includegraphics[width=1in,height=1.25in,clip,keepaspectratio]{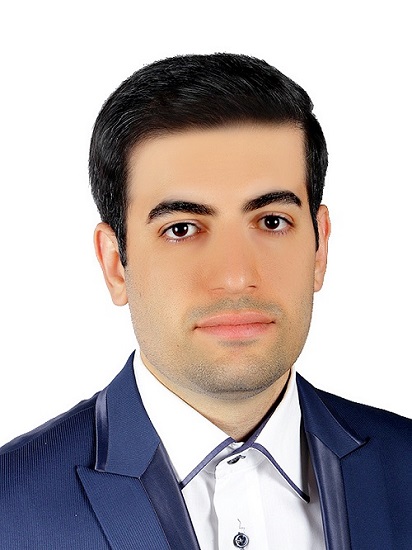}}]{Rahman Saadat Yeganeh}
received his Ph.D. degree in Electrical Engineering (Telecommunication Systems) from Isfahan University of Technology (IUT) in 2024. He is currently a Postdoctoral Researcher in Telecommunication Systems at Sharif University of Technology (SUT), Tehran, Iran. Beyond academia, Dr. Saadat is actively involved in the research and design of various telecommunication systems. His research interests include wireless communication systems with a particular focus on 6G cellular networks and IoT, as well as satellite communications, symbiotic radio, reconfigurable intelligent surfaces (RIS), nonorthogonal multiple access (NOMA), and signal processing.
\end{IEEEbiography}

\begin{IEEEbiography}[{\includegraphics[width=1in,height=1.25in,clip,keepaspectratio]{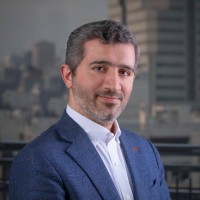}}]{Hamid Behroozi}
(Member, IEEE) received the B.Sc. degree in electrical engineering from the University of Tehran, Tehran, Iran, in 2000, the M.Sc. degree in electrical engineering from the Sharif University of Technology, Tehran, in 2003, and the Ph.D. degree in electrical engineering from Concordia University, Montreal, QC, Canada, in 2007. From 2007 to 2010, he was a Postdoctoral Fellow with the Department of Mathematics and Statistics, Queen’s University, Kingston, ON, Canada. He is currently an Associate Professor with the Department of Electrical Engineering, Sharif University of Technology. His research interests include information theory, joint source-channel coding, artificial intelligence in signal processing and data science, and cooperative communications. He was the recipient of several academic awards, including the Ontario Postdoctoral Fellowship awarded by the Ontario Ministry of Research and Innovation, the Quebec Doctoral Research Scholarship awarded by the Government of Quebec, the Hydro Quebec Graduate Award, and the Concordia University Graduate Fellowship
\end{IEEEbiography}

\begin{IEEEbiography}[{\includegraphics[width=1in,height=1.25in,clip,keepaspectratio]{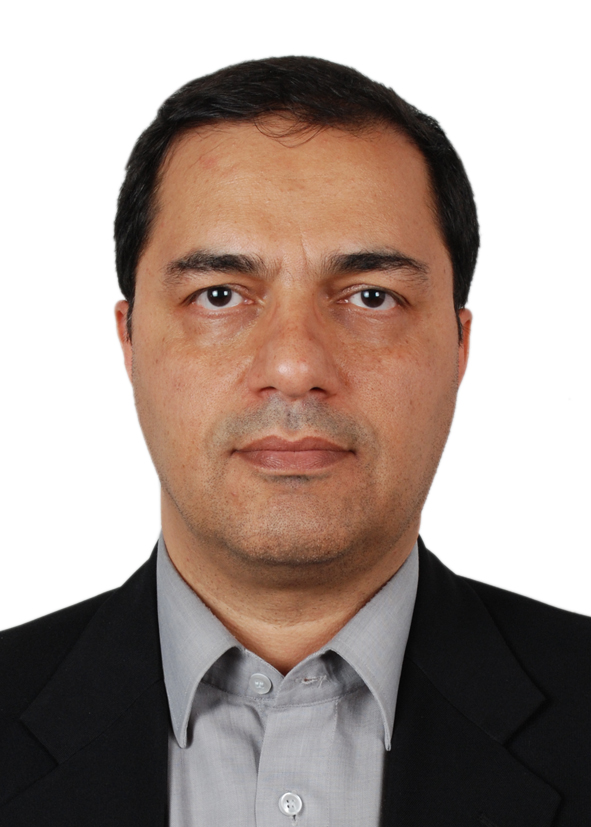}}]{Mohammad Javad Omidi}
received the Ph.D. degree from the University of Toronto in 1998. He has extensive industry experience in Canada, specializing in the design of broadband communication systems. He has held several prominent academic and administrative positions with the Isfahan University of Technology, Iran, including as a Professor with the Department of Electrical and Computer Engineering (ECE), the Chair of the IT Center, the Chair of the ECE Department, and the Vice President for Research and Technology.

He currently serves as a Professor with the ECE Department, Kuwait College of Science and Technology. Beyond academia, he has played a significant role in fostering innovation and entrepreneurship. He has extensive experience in managing science parks, supporting technology start-ups, and mentoring university graduates in their entrepreneurial ventures. His research focuses on wireless communications, digital communication systems, and cognitive radio systems. He has authored numerous publications and holds six U.S. patents along with four international patents in these areas.
\end{IEEEbiography}
\begin{IEEEbiography}[{\includegraphics[width=1in,height=1.25in,clip,keepaspectratio]{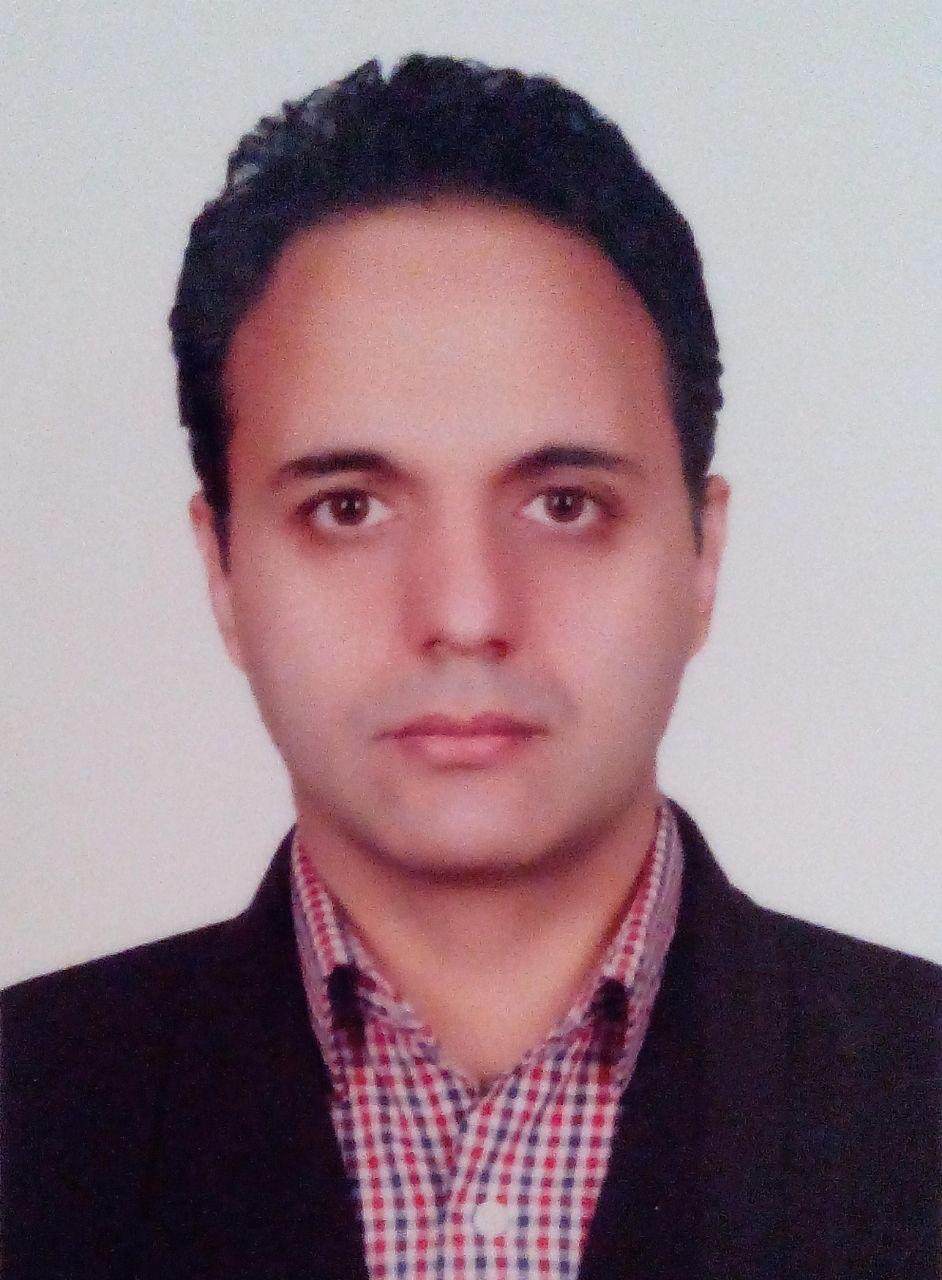}}]{Mohammad Robat Mili}
received the Ph.D. degree in electrical and electronic engineering from the University of Manchester, U.K., in 2012. He held postdoctoral research positions at the Department of Telecommunications and Information Processing, Ghent University, Belgium and the Department of Electrical Engineering, Sharif University of Technology, Iran. His main research interests are in the area of design and analysis of wireless communication networks with particular focus on 5G and 6G cellular networks using mathematical methods such as optimization theory, game theory, and machine learning
\end{IEEEbiography}

\begin{IEEEbiography}
[{\includegraphics[width=1in,height=1.25in,clip,keepaspectratio]{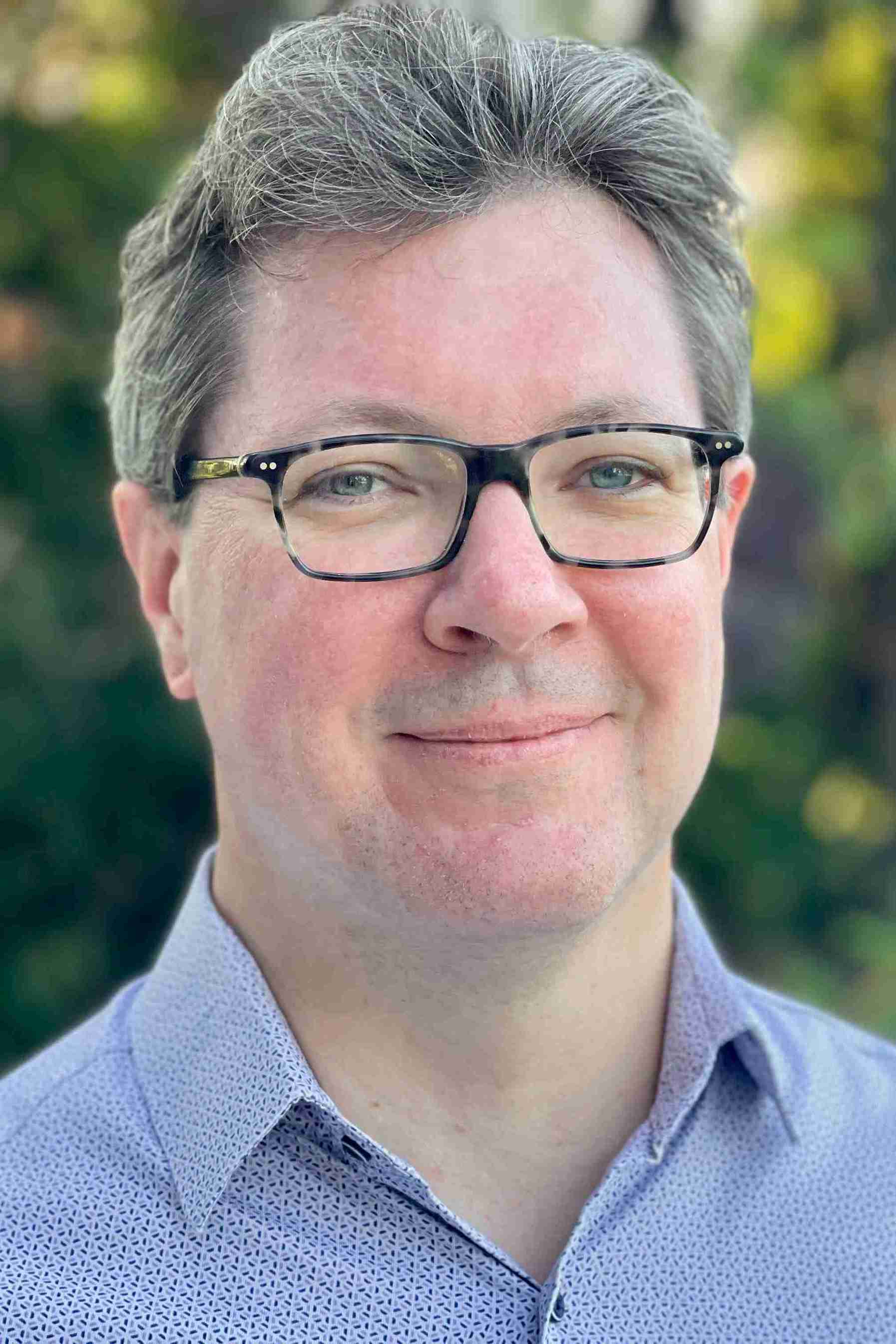}}]{ Eduard Axel Jorswieck}
is full professor with the Faculty of Electrical Engineering, Information Technology, Physics of TU Braunschweig. He is Dean of the Faculty for Electrical Engineering, Information Technology, Physics and the Managing Director of the Institute for Communications Technology at TU Braunschweig, Germany. From 2008 until 2019, he was the head of the Chair for Communications Theory and Full Professor at TU Dresden, Germany. He is IEEE Fellow. His general interests are in signal processing for communications and networking, applied information theory and communication theory. He has published more than 200 journal articles, 19 book chapters, one book, four monographs, and some 350 conference papers. He was a co-recipient of the IEEE Signal Processing Society Best Paper Award 2006. His colleagues and he were also recipients of the Best Paper Awards and the Best Student Paper Awards from the IEEE CAMSAP 2011, IEEE WCSP 2012, IEEE SPAWC 2012, IEEE ICUFN 2018, PETS 2019, and ISWCS 2019, and IEEE ICC 2024. 

Since 2017, he has been the Editor-in-Chief of the EURASIP Journal on Wireless Communications and Networking. Since 2024, he has been an Editor for IEEE TRANSACTIONS ON INFORMATION THEORY. He was on the editorial boards of the IEEE SIGNAL PROCESSING LETTERS, IEEE TRANSACTIONS ON SIGNAL PROCESSING, the IEEE TRANSACTIONS ON WIRELESS COMMUNICATIONS, IEEE TRANSACTIONS ON INFORMATION FORENSICS AND SECURITY, and IEEE TRANSACTIONS ON COMMUNICATIONS. He received the 2019 outstanding editorial board award from the IEEE Transactions on Information Forensics and Security. 

Eduard has been general, technical program chair and organizing committee member of many international conferences of the last 20 years. Most recently, he was general co-chair of the ITG Workshop Smart Antennas / Conference on Systems, Communications, and Coding in 2023 in Braunschweig. He was tutorial chair for IEEE Globecom 2025 and is TPC Co-Chair of IEEE SPAWC 2026. He has contributed as guest editor to several special issues in IEEE Journals. Most recently, he was guest editor in the IEEE Journal on Selected Areas in Communications for Advanced Optimization Theory and Algorithms for Next Generation Wireless Communication Networks (2024) and in IEEE Journal on Selected Topics in Signal Processing on Distributed Signal Processing for Extremely Large-Scale Antenna Array Systems (2025). 
\end{IEEEbiography}

\begin{IEEEbiography}
[{\includegraphics[width=1in,height=1.25in,clip,keepaspectratio]{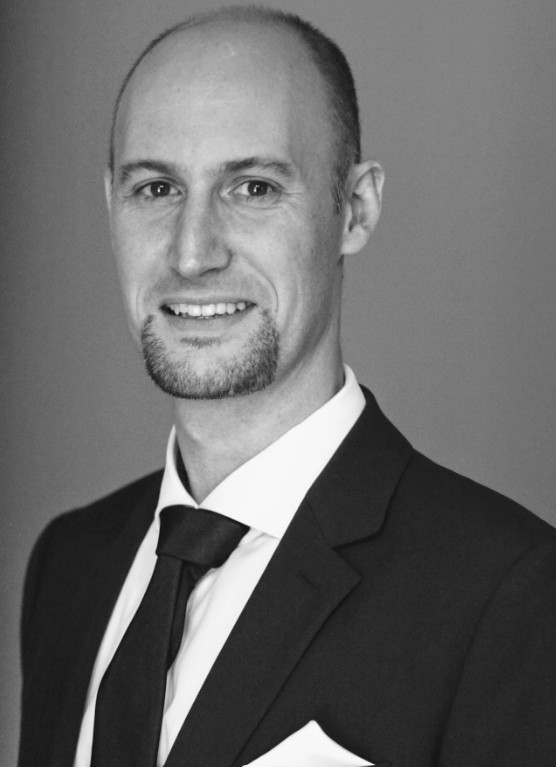}}]{Symeon Chatzinotas} 
is currently Full Professor / Chief Scientist I and Head of the research group SIGCOM in the Interdisciplinary Centre for Security, Reliability and Trust, University of Luxembourg. In parallel, he is an Adjunct Professor in the Department of Electronic Systems, Norwegian University of Science and Technology, an Eminent Scholar of the Kyung Hee University, Korea and a Collaborating Scholar of the Institute of Informatics \& Telecommunications, National Center for Scientific Research “Demokritos”. 

In the past, he has been a Visiting Professor at EPFL, Switzerland and University of Parma, Italy and contributed in numerous R\&D projects for the Institute of Telematics and Informatics, Center of Research and Technology Hellas and Mobile Communications Research Group, Center of Communication Systems Research, University of Surrey.

He has received the M.Eng. in Telecommunications from Aristotle University of Thessaloniki, Greece and the M.Sc. and Ph.D. in Electronic Engineering from University of Surrey, UK in 2003, 2006 and 2009 respectively. 

He has authored more than 800 technical papers in refereed international journals, conferences and scientific books and has received numerous awards and recognitions, including the IEEE Fellowship, IEEE Distinguished Contributions Award and IEEE Harry Rowe Mimno Award. He has served in the editorial board of npj Wireless Technology, IEEE Transactions on Communications, IEEE Open Journal of Vehicular Technology and the International Journal of Satellite Communications and Networking.
\end{IEEEbiography}

\end{document}